\documentclass[usenatbib]{mn2e}
\usepackage{graphicx}

\newcommand{\cmt}   {~cm$^{-3}$}

\newcommand{\hcop} {HCO$^+$}

\newcommand{\cdh}  {C$_2$H}
\newcommand{\cqh}  {C$_4$H}
\newcommand{\cthd} {C$_3$H$_2$}
\newcommand{\hctn} {HC$_3$N}
\newcommand{\nh}   {NH$_3$}
\newcommand{\nthp} {N$_2$H$^+$}
\newcommand{\hcsp} {HCS$^+$}

\begin{document}

\title{Time-dependent models of dense PDRs with complex molecules}

\author[O.\ Morata \& E.\ Herbst]{
Oscar Morata$^{1}$\thanks{Currently at the Institute of Astronomy and
  Astrophysics, Academia Sinica \& Department of Earth Sciences, National
  Taiwan Normal University, 88, Sec. 4, Ting-Chou Road, Taipei 11677, Taiwan,
  ROC} and Eric Herbst$^{2}$ \\
$^{1}$ Department of Physics, The Ohio State University, Columbus, OH 43210,
USA\\ 
$^{2}$ Departments of Physics, Astronomy, and Chemistry, The Ohio State
University, Columbus, OH 43210, USA\\}

\maketitle

\begin{abstract}
We present a study of the chemistry of a dense photon-dominated region (PDR)
using a time-dependent chemical model.  Our major interest is to study the
spatial distribution of complex molecules such as hydrocarbons and
cyanopolyynes in the cool dense material bordering regions where star
formation has taken place. Our standard model uses a homogeneous cloud of
density $2\times10^4$\cmt\ and temperature $T=40$ K, which is irradiated by a
far-ultraviolet radiation field of intermediate intensity, given by
$\chi=100$. We find that over a range of times unsaturated hydrocarbons (e.g.,
\cdh, \cqh, \cthd) have relatively high fractional abundances in the more
external layers of the PDR, whereas their abundances in the innermost layers
are several orders of magnitudes lower. On the other hand, molecules that are
typical of late-time chemistry are usually more abundant in the inner parts of
the PDR. We also present results for models with different density,
temperature, intensity of the radiation field and initial fractional
abundances. Our results are compared with both high- and moderate-angular
resolution observations of the Horsehead nebula.  Our standard model is
partially successful in reproducing the observations. Additional models run
with different physical parameters are able to reproduce the abundance of many
of the observed molecules, but we do not find a single model that fits all the
observations at the same time.  We discuss the suitability of a time-dependent
model of a dense PDR such as ours as an estimator of the age of a PDR,
provided that enough observational data exist.
\end{abstract}

\begin{keywords}
ISM: molecules -- ISM: abundances -- Stars: formation -- astrochemistry --
molecular processes
\end{keywords}

\section{Introduction}

Far-ultraviolet (FUV) photons (6 eV$<h\nu<13.6$ eV) can drastically affect the
structure, chemistry, thermal balance, and evolution of portions of the
interstellar medium, even if only the mean galactic interstellar radiation
field (ISFR) is present. For this reason, the simplest chemical models of
dense interstellar clouds should include a treatment of the radiative transfer
of this radiation through the outer and intermediate layers of material,
because the field affects observations that also sample the interior gas.
Indeed, most of the molecular gas and certainly all the atomic gas of the
Galaxy is found in regions where the FUV photon flux is several times that of
the mean ISRF \citep{Hollenbach97}.  Photon-dominated regions (PDRs) are
special regions of neutral gas exposed to intense fluxes of FUV radiation,
which dominate the physical (heating) and chemical processes that take place
in their interiors. PDRs can be observed in very different environments,
including dense molecular clouds such as the Orion Bar or the Horsehead
nebula, planetary nebulae, winds of red giants and AGB stars, the interstellar
medium of starburst galaxies such as M82, low metallicity dwarf galaxies, etc.
PDRs are studied through atomic spectral lines such as [C\,{\small II}] at
158$\mu$m, [O\,{\small I}] at 63 and 146 $\mu$m, [C\,{\small I}] at 370 and 609
$\mu$m; many H$_2$O lines; millimetre and sub-millimetre rotational
transitions of molecules such as CO, $^{13}$CO, or C$^{18}$O; ro-vibrational
and pure rotational lines of H$_2$; and many broad mid-IR features attributed
to PAHs.

Physical conditions of PDRs are thus very diverse.  They can be diffuse, with
gas density $n\sim 10-100$~cm$^{-3}$, or dense, with $n > 10^4$ cm$^{-3}$,
while the incident FUV flux may range from the ISRF to $10^6$ times the ISRF
very near to an O star.  An important characteristic of PDRs is that they
have a layered structure, as a result of the interaction of the radiation with
the gas and dust. Typically, they contain an outer layer of partially ionised
gas, where H is atomic and the predominant form of carbon is C\,{\small II}; a
layer of neutral gas, where H$_2$ is predominant and carbon is present mainly
as C\,{\small I}; and a molecular layer, where CO is the main form of
carbon. The ratio between the flux of the incident radiation field and the gas
density determines the size of these layers. The distinct layers should be
observable if the PDR is seen edge-on, because then the geometry of the cloud
is more apparent. One of the main difficulties of the study of PDRs is,
however, the determination of the geometry of the object.  The gas temperature
is also dependent on the strength of the radiation field and the position
inside the cloud, and can range from 10-100 K in the molecular layer to more
than 10$^3$ K in the surface layers.

\begin{table}
\caption{Standard initial (dark cloud) fractional abundances with respect to
$n_{\rm H}$}
\label{iniabund}
\begin{center}
\begin{tabular}{llll}
\noalign{\smallskip}\hline\noalign{\smallskip}
H$_2$  & 0.5                  &
   C$_2$H & $1.0 \times 10^{-8}$ \\
H      & $7.5 \times 10^{-5}$ &
   CO$_2$ & $1.3 \times 10^{-8}$ \\
He     & 0.14                 &
   H$_2$O & $3.5 \times 10^{-8}$ \\
C      & $2.8 \times 10^{-8}$ &
   HCN    & $1.0 \times 10^{-8}$ \\
O      & $1.0 \times 10^{-4}$ &
   HNC    & $1.0 \times 10^{-8}$ \\
N      & $1.3 \times 10^{-5}$ &
   NH$_3$ & $1.0 \times 10^{-8}$ \\
S      & $7.2 \times 10^{-8}$ &
   SO$_2$ & $5.0 \times 10^{-10}$ \\
Si     & $7.8 \times 10^{-9}$ &
   C$_3$H & $5.0 \times 10^{-9}$ \\
Cl     & $4.0 \times 10^{-9}$ &
   C$_4$H & $4.5 \times 10^{-8}$ \\
Fe     & $3.9 \times 10^{-10}$ &
   C$_3$H$_2$ & $5.0 \times 10^{-9}$ \\
Mg     & $1.9 \times 10^{-9}$ &
   HC$_3$N & $1.0 \times 10^{-8}$ \\
Na     & $4.7 \times 10^{-10}$ &
   C$^+$  &  $4.7 \times 10^{-9}$ \\
P      & $3.0 \times 10^{-9}$ &
   H$^+$  &  $4.2 \times 10^{-10}$ \\
CH     & $1.0 \times 10^{-8}$ &
   He$^+$  & $3.5 \times 10^{-10}$ \\
CN     & $2.5 \times 10^{-9}$ &
   Fe$^+$ &  $2.6 \times 10^{-9}$ \\
CO     & $7.3 \times 10^{-5}$ &
   Mg$^+$ &  $5.1 \times 10^{-9}$ \\
CS     & $2.0 \times 10^{-9}$ &
   Na$^+$ &  $1.5 \times 10^{-9}$ \\
N$_2$  & $4.2 \times 10^{-6}$ &
   S$^+$  &  $1.2 \times 10^{-9}$ \\
NO     & $1.5 \times 10^{-8}$ &
   Si$^+$ &  $2.5 \times 10^{-10}$ \\
O$_2$  & $8.1 \times 10^{-8}$ &
   H$_3^+$ & $1.4 \times 10^{-9}$ \\
OH     & $1.0 \times 10^{-7}$ &
   HCO$^+$ & $4.0 \times 10^{-9}$ \\
S$_2$  & $1.8 \times 10^{-9}$ &
   HCS$^+$ & $2.0 \times 10^{-10}$ \\
SO     & $1.0 \times 10^{-9}$ &
   N$_2$H$^+$ & $2.0 \times 10^{-10}$ \\
\noalign{\smallskip}\hline\noalign{\smallskip}
\end{tabular}
\end{center}
\end{table}

Dense PDRs are of particular interest in our attempt to understand the
process of star formation. Newly formed stars produce high intensity photon
fields deep inside clouds that lead to the formation of H\,{\small II} regions
and, further removed from the exciting stars, PDRs. The gas-phase chemistry
that occurs under these conditions is in most layers greatly different from
the chemistry found in quiescent regions. The abundances of some atoms or
molecules are greatly enhanced by the presence of the FUV field, while other
species are destroyed by the radiation. For instance, the abundances of
several hydrocarbons (C$_2$H, c-\cthd, C$_4$H, C$_6$H) have been found to be
almost as high in relatively unshielded regions as in dark clouds
\citep{Fuente03, Teyssier04}. However, some of these high abundances are
difficult to explain with normal gas-phase chemical models and a link between
small hydrocarbons and PAHs has been proposed \citep{Teyssier04}.

In this paper, we study the chemistry of a dense PDR created after a star has
been formed inside a dark cloud. We are particularly interested in the spatial
distributions of hydrocarbons and other complex molecules, and whether or not
time-dependent chemistry can explain some of the discrepancies found between
the observed abundances and the results of steady-state models
\citep{Teyssier04,Pety05} for the Horsehead nebula, as well as providing us
with some estimate of the age of this PDR. Another use of time-dependent PDR
chemistry is described by \citet{Bell05}. The structure of the remainder of
the paper is as follows: in Section~\ref{sct_model}, we describe the
time-dependent chemical model used.  Our results obtained with our standard
model are then considered in Section~\ref{sct_results}, while in
Section~\ref{sct_disc} we discuss the dependence of our results on assorted
physical parameters.  The agreement between model results and observations of
the Horsehead nebula is probed as a function of time in
Section~\ref{hhnebula}.  Finally, Section~\ref{sectsum} contains a summary of
the paper.

\section{Model}
\label{sct_model}

\begin{figure}
 \begin{center}
   \includegraphics[width=8.4cm]{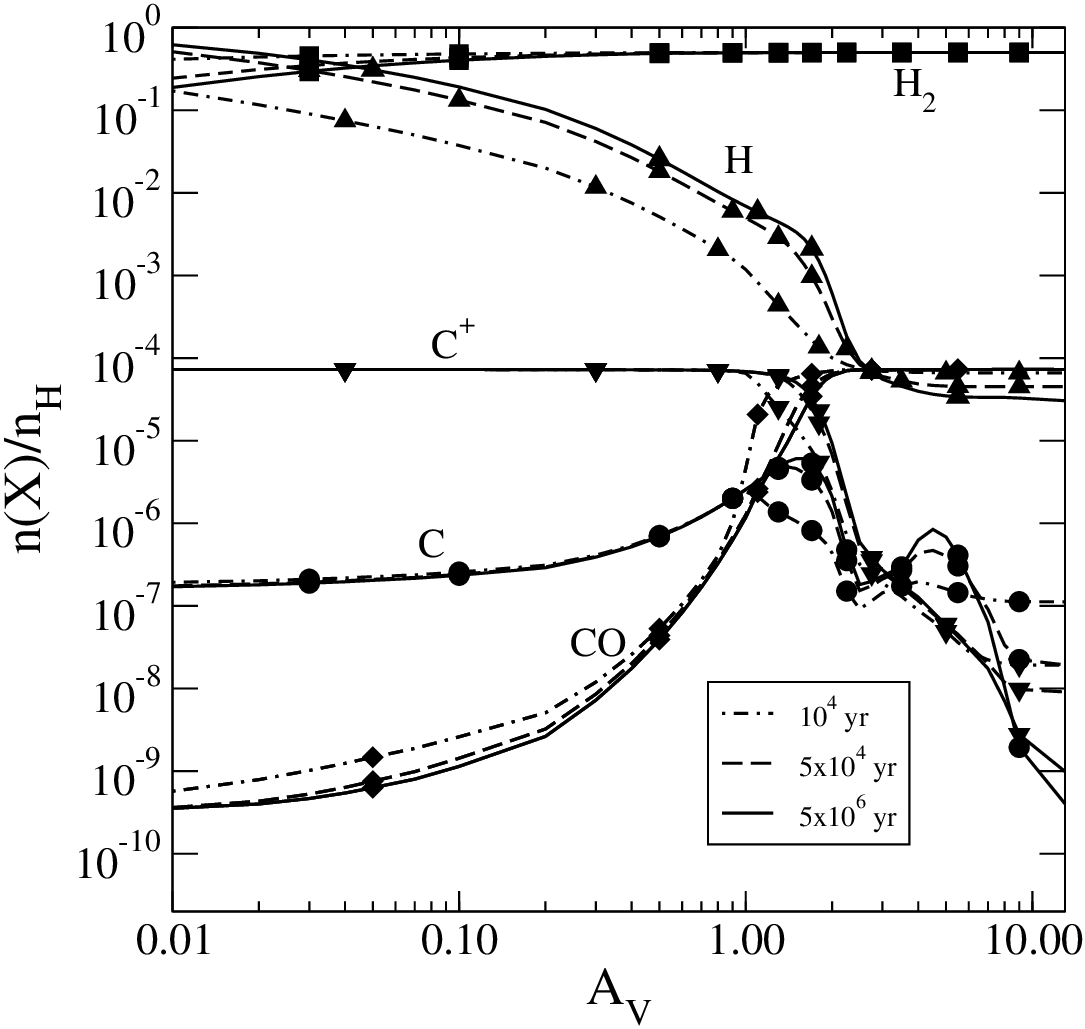}
   \caption{Fractional abundances of H (filled triangles), H$_2$ (filled
   squares), C (filled circles), C$^+$ (filled triangles pointed downward),
   and CO (filled diamonds) with respect to $n_{\rm H}$ as a function of A$_V$
   for our standard model. Three different times are shown: $10^4$ yr
   (dashed-dotted lines), $4\times10^4$ yr (dashed lines), and $5\times10^6$ yr
   (solid lines). The solid lines correspond to steady-state abundances. }
   \label{bigones}
 \end{center}
\end{figure}

In our model, based on \citet{Lee96}, we assume a semi-infinite plane-parallel
cloud divided into $n_i$ slabs, on one side of which impinges an ultraviolet
radiation field, defined in our case in units of the \citet{Draine78} standard
radiation field scaled by a multiplicative factor $\chi$. The visual
extinction, A$_V$, and the total column density, N$_{\rm H}$, are related by
A$_V= 6.289\times10^{-22}$ N$_{\rm H}$ \citep{Wagenblast89}. The model treats
radiative transfer and, in particular, takes into account the self- and
cross-shielding due to H$_2$ and CO. We use the \citet{DraineBert96}
analytical approximation formula to calculate the H$_2$ photo-dissociation
rate ($b$ = 1 km s$^{-1}$), which we have found to be in closer
agreement with the results of the `exact' treatment of H$_2$ self-shielding
\citep*{LePetit02} than the method presented in \citet{Lee96}. However, the
larger differences are only found for small values of the visual extinction,
A$_V < 0.01$, which are not the main source of interest of the present paper.
For the CO photodissociation rate, we use the same method as in \citet{Lee96}.

\begin{figure}
 \begin{center}
   \includegraphics[width=8.4cm]{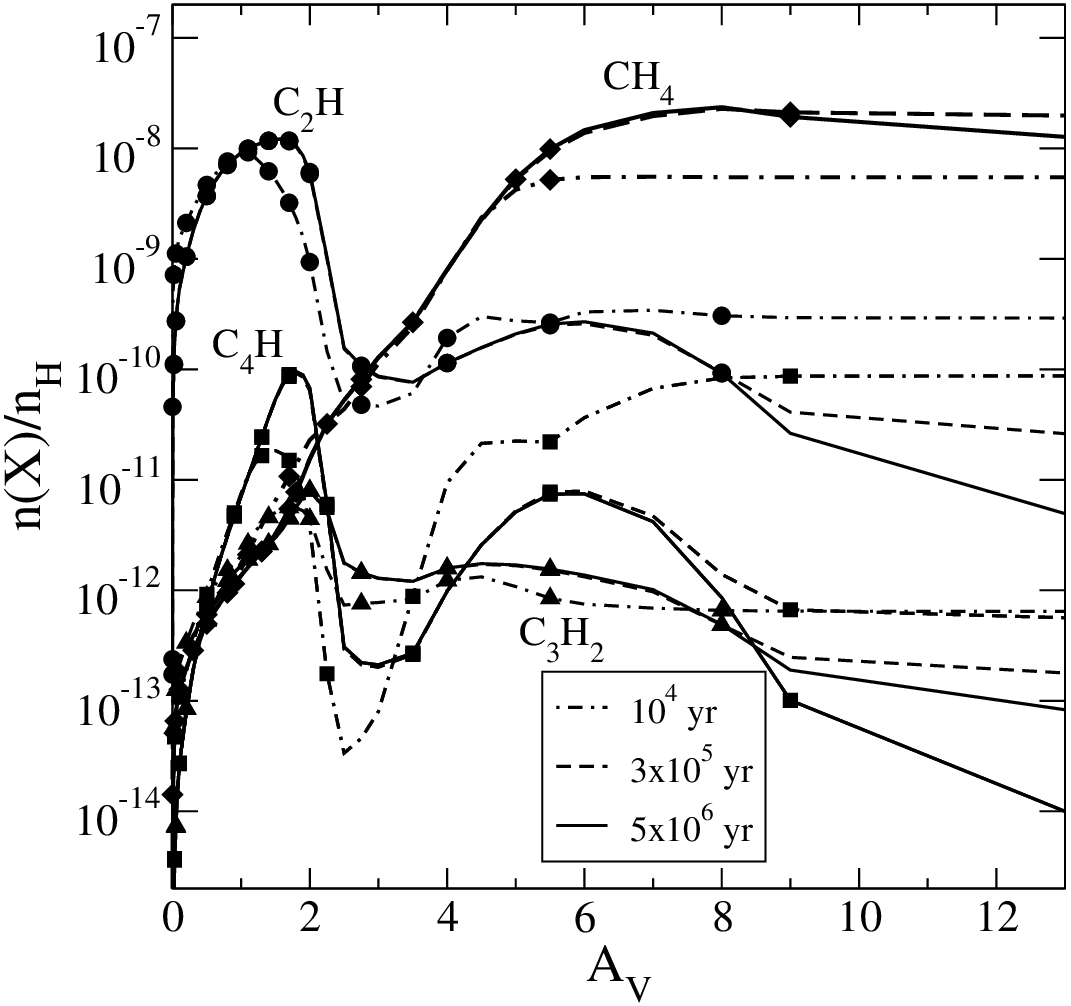}
   \caption{Fractional abundances of the molecules CH$_4$ (filled diamonds),
   \cdh\ (filled circles), \cqh\ (filled squares), and \cthd\ (filled
   triangles) with respect to $n_{\rm H}$ as a function of A$_V$ for our
   standard model. Three different times are shown: $10^4$ yr
   (dashed-dotted line), $3\times10^5$ yr (dashed line), and $5\times10^6$
   yr (solid line). The solid lines correspond to steady-state
   abundances. }
   \label{carbon1}
 \end{center}
\end{figure}

For our standard model, we have assumed a homogeneous cloud with physical
conditions similar to those observed in the Horsehead nebula (see
Section~\ref{hhnebula}), namely a density $n_{\rm{H}}=2\times10^4$\cmt\ and a
temperature $T=40$~K. The cosmic-ray ionisation rate, $\zeta$, is assumed to
be $1.3\times10^{-17}$~s$^{-1}$. We use a value for the scaling factor of the
radiation field of $\chi=100$, which is similar to the one measured in the
Horsehead nebula \citep{Zhou93,Abergel03}. In our standard model, the cloud is
divided into 42 slabs of differing length, defined in a way so as to have a
smooth distribution of visual extinction points from A$_V=5\times10^{-4}$ to
A$_V=13$, but good coverage of the positions where the fractional abundances
of most of the species show steeper slopes.  The time-dependent gas-phase
chemistry is solved slab-wise, starting from the outermost slab and moving
successively inward. The model solves a system of ordinary differential
equations with the Gear method for each slab using the H$_2$ and CO
photodissociation rates from the {\em (i-1)th} slab as the input values for
the {\em i-th} slab. For the chemistry, we use the \texttt{osu.2003}
\citep*{Smith04} gas-phase chemical network, which contains 419 species and
4103 reactions.  This relatively old version of the osu network was used for a
variety of reasons.  First, it was the model with which the original
calculations were begun.  During the calculations, several newer versions of
the network \texttt{(osu.2005)} were prepared, but required modifications and
corrections, which were not fully resolved until the summer of 2007.  To
determine if the newest version of the osu code (\texttt{osu\_03\_2008}; see
http://www.physics.ohio-state.edu/$\sim$eric/) leads to any significant
changes, we have run a new version of our standard model with this network.
The differences are discussed in Appendix A.

We assume that our dense PDR is formed as a result of the process of star
formation inside a dense cloud, when radiation from a new star begins to
affect the chemistry of the surrounding gas. Thus, we assume that, at $t=0$,
the gas has already undergone chemical evolution during $\sim10^5$ yr under
dense conditions: a density $n_{\rm{H}}=2\times10^4$\cmt\ and $T=10$ K. The
initial abundances are those of a typical dark cloud (see
Table~\ref{iniabund}) as derived from the values measured for smaller species
(composed of up to five atoms) in TMC-1CP \citep{Smith04}. For those smaller
species where there is no available observed value, we adopted the fractional
abundances calculated by \citet{Smith04} at 10$^5$ yr. In our model, we
maintain the standard depleted cosmic abundance ratios C/H$=7.3\times10^{-5}$
and O/H$=1.76\times10^{-4}$. Since the total C/H abundance measured in TMC-1P
is a factor $\sim2$ less than the standard cosmic ratio, we have assumed that
the remaining C is locked up in CO, with the initial abundance shown in
Table~\ref{iniabund}. The initial abundance of atomic O is then chosen so that
the overall standard depleted O/H abundance is used. We discuss in
Section~\ref{sectelemental} other possible results using the lower initial
abundances of CO.

\begin{figure}
 \begin{center}
   \includegraphics[width=8.4cm]{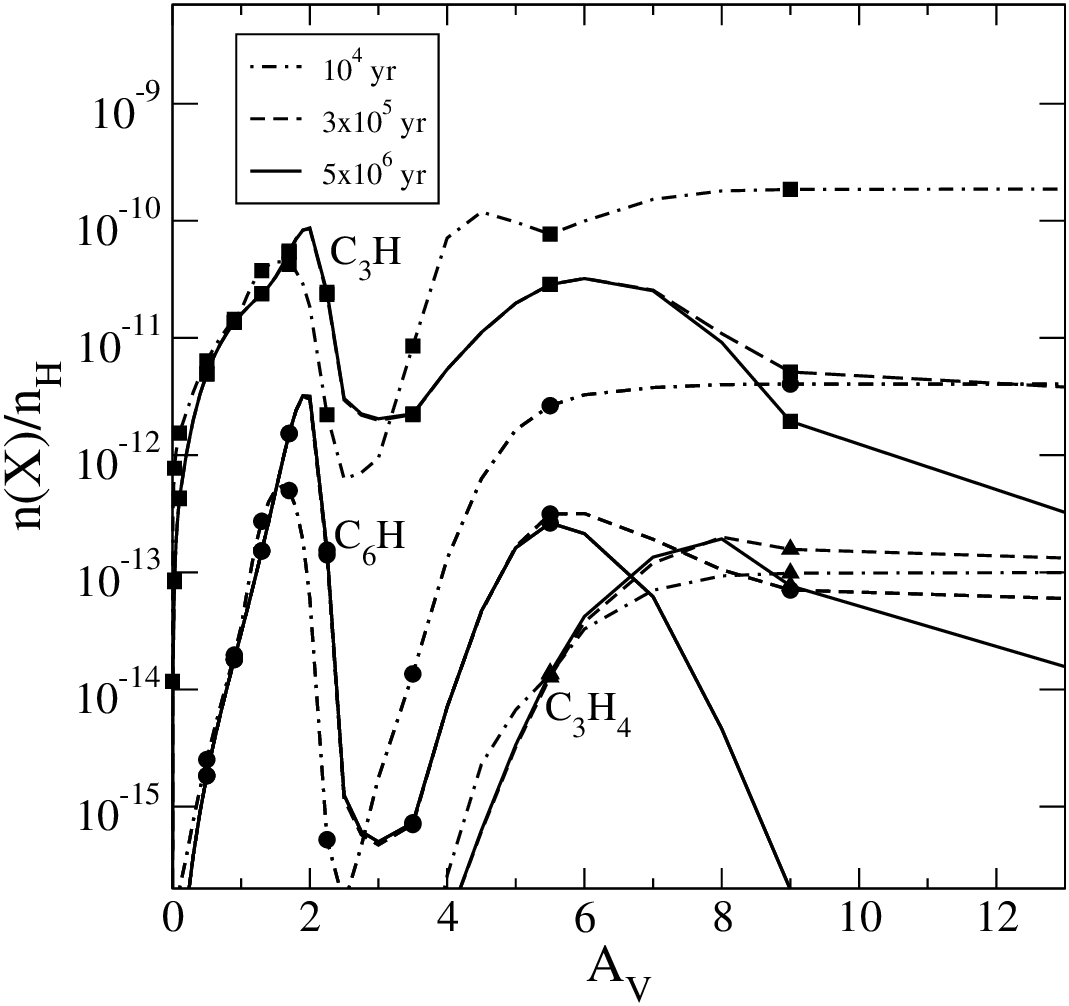}
   \caption{Same as Fig.~\ref{carbon1}, but for the molecules C$_3$H (filled
   squares), C$_6$H (filled circles), and C$_3$H$_4$ (filled triangles).}
   \label{carbon2}
 \end{center}
\end{figure}

The $t=0$ values listed in Table~\ref{iniabund} are more general than those
found for TMC-1, where a much more complex molecule development has already
occurred, which appears to be unique.

\begin{table*}
 \caption[]{Peak fractional abundances and A$_V$ for selected
hydrocarbons in the standard model}
 \label{carbonpeaks}
\begin{center}
 \begin{tabular}{lclrclrcll}
\noalign{\smallskip}\hline\noalign{\smallskip}
Species & & \multicolumn{2}{c}{$t=10^4$ yr} & &
\multicolumn{2}{c}{$t=3\times10^5$ yr} & &
\multicolumn{2}{c}{$t=5\times10^6$ yr} \\
&& n(X)/n$_{\rm H}$ & A$_V$ && n(X)/n$_{\rm H}$ & A$_V$ && n(X)/n$_{\rm H}$ &
A$_V$ \\
\noalign{\smallskip}\hline\noalign{\smallskip}
CH$_4$ && $5.5\times10^{-9}$ & 7.0 && $2.3\times10^{-8}$ & 8.0 && $2.4\times10^{-8}$ & 8.0\\
CH && $1.6\times10^{-8}$ & 1.1 && $1.9\times10^{-8}$ & 1.6 && $1.9\times10^{-8}$ & 1.6\\
C$_2$H && $9.3\times10^{-9}$ & 1.0 && $1.2\times10^{-8}$ & 1.6 && $1.2\times10^{-8}$ & 1.6\\
CH$_2$ && $1.4\times10^{-8}$ & 1.1 && $7.3\times10^{-9}$ & 1.8 && $7.3\times10^{-9}$ & 1.8\\
C$_2$H$_2$ && $4.2\times10^{-9}$ &13.0 && $1.5\times10^{-9}$ & 8.0 && $1.5\times10^{-9}$ & 8.0\\
CH$_3$ && $2.3\times10^{-10}$ & 2.0 && $2.4\times10^{-10}$ & 2.2 && $2.4\times10^{-10}$ & 2.2\\
C$_4$H && $8.7\times10^{-11}$ &13.0 && $9.2\times10^{-11}$ & 1.8 && $9.5\times10^{-11}$ & 1.8\\
C$_3$H && $1.9\times10^{-10}$ &13.0 && $8.6\times10^{-11}$ & 2.0 && $8.7\times10^{-11}$ & 2.0\\
C$_3$H$_2$ && $1.0\times10^{-9}$ &13.0 && $6.2\times10^{-11}$ & 1.9 && $6.3\times10^{-11}$ & 1.9\\
C$_5$H && $1.0\times10^{-11}$ &13.0 && $9.6\times10^{-12}$ & 1.9 && $9.9\times10^{-12}$ & 2.0\\
C$_2$H$_3$ && $5.6\times10^{-12}$ & 1.7 && $8.0\times10^{-12}$ & 2.0 && $8.0\times10^{-12}$ & 2.0\\
C$_6$H && $4.0\times10^{-12}$ &13.0 && $3.1\times10^{-12}$ & 1.9 && $3.2\times10^{-12}$ & 1.9\\
C$_2$H$_4$ && $4.0\times10^{-13}$ &13.0 && $1.8\times10^{-12}$ & 9.0 && $1.9\times10^{-12}$ & 8.0\\
C$_4$H$_2$ && $2.5\times10^{-10}$ &13.0 && $2.1\times10^{-12}$ & 8.0 && $1.8\times10^{-12}$ & 7.0\\
C$_8$H && $2.0\times10^{-13}$ & 9.0 && $2.7\times10^{-13}$ & 2.0 && $2.7\times10^{-13}$ & 2.0\\
C$_3$H$_4$ && $1.0\times10^{-13}$ &13.0 && $2.0\times10^{-13}$ & 8.0 && $1.9\times10^{-13}$ & 8.0\\
C$_3$H$_3$ && $1.3\times10^{-14}$ &13.0 && $3.3\times10^{-14}$ & 7.0 && $5.8\times10^{-14}$ & 7.0\\
C$_5$H$_2$ && $5.2\times10^{-12}$ &13.0 && $3.2\times10^{-14}$ & 7.0 && $3.5\times10^{-14}$ & 7.0\\
C$_7$H && $4.4\times10^{-13}$ &13.0 && $4.6\times10^{-14}$ & 6.0 && $2.7\times10^{-14}$ & 5.5\\
C$_4$H$_3$ && $1.3\times10^{-13}$ &13.0 && $2.3\times10^{-14}$ & 7.0 && $2.5\times10^{-14}$ & 7.0\\
CH$_3$C$_4$H && $1.1\times10^{-13}$ &13.0 && $2.1\times10^{-14}$ & 7.0 && $2.2\times10^{-14}$ & 7.0\\
C$_6$H$_2$ && $5.4\times10^{-12}$ &13.0 && $2.1\times10^{-14}$ &13.0 && $1.6\times10^{-14}$ & 7.0\\
C$_6$H$_6$ && $4.7\times10^{-14}$ &13.0 && $1.2\times10^{-14}$ & 7.0 && $2.1\times10^{-15}$ & 7.0\\
\noalign{\smallskip}\hline\noalign{\smallskip}
\end{tabular}
\end{center}
\end{table*}

\section{Results}
\label{sct_results}

Fig.~\ref{bigones} shows  the fractional abundance profiles of  H, H$_2$, C,
C$^+$, and CO obtained with our  standard model at selected times up to steady
state ($ 5\times 10^{6}$ yr).  Such a profile at $t = 0$ would be totally flat
at  the  value  given  in  Table  1.   Here  the  three  layers  described  in
Sect.~\ref{sct_model}  can be  distinguished. The  crossing point  between the
fractional abundances of H and H$_2$  moves inwards with time, and it is found
at A$_V\sim0.04$ at steady-state. The  low value for this visual extinction is
somewhat artificial; proper  inclusion of factors such as  a lower density and
higher temperature at the cloud edge and a less efficient surface formation of
H$_{2}$ would increase the extinction at crossing.  Further into the cloud, CO
becomes  the most  abundant carbon-bearing  species at  A$_V\sim1.7$,  while C
becomes more abundant  than C$^{+}$ at visual extinctions  between 3 and 8.5.

\subsection{Hydrocarbons}
\label{secthydrocarbons}

Figs~\ref{carbon1} and \ref{carbon2} show the fractional abundances of
several hydrocarbon molecules as a function of A$_V$ for our standard model at
three different times: $t = 10^4$, $3\times10^5$, and $5\times10^6$ yr.  The
first two times correspond, respectively, to an early time in the evolution
when only the outermost layers are approaching steady state, and a time when
steady-state is virtually reached in the outer to intermediate regions of the
PDR (A$_V<3-5$).  In addition to the figures, Table~\ref{carbonpeaks} shows
the peak fractional abundances and corresponding visual extinctions of most
hydrocarbons present in our model at the three different times.  In this and
subsequent tables of fractional abundances, the order of presentation is in
terms of decreasing steady-state peak fractional abundance.

\begin{figure}
 \begin{center}
   \includegraphics[width=8.4cm]{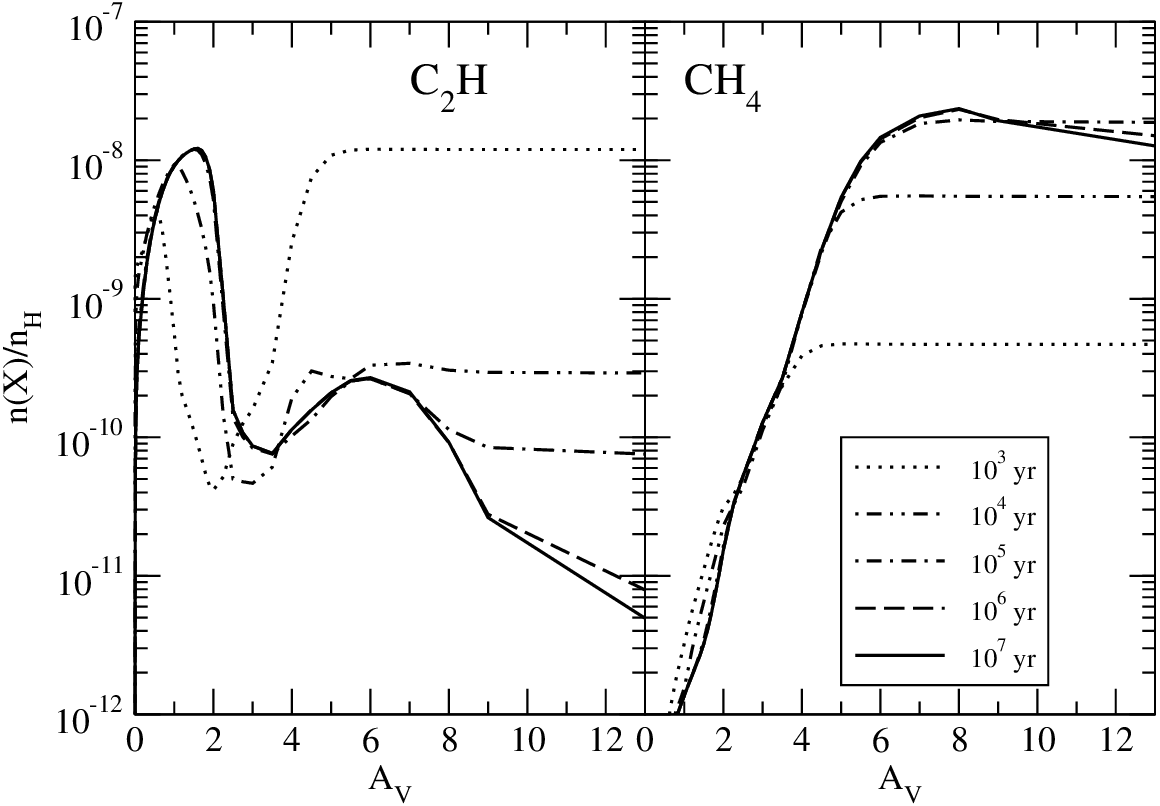}
   \caption{Fractional abundances with respect to $n_{\rm H}$ for the molecules
   \cdh\ (left panel) and CH$_4$ (right panel) as a function of A$_V$ in our
   standard model for five different times: $10^3$, $10^4$, $10^5$, $10^6$, and
   $10^7$ yr. The solid lines represent steady-state
   abundances. }
   \label{earlytimes}
 \end{center}
\end{figure}

The evolution of the fractional abundances of a hydrocarbon depends upon
whether the molecule is present initially in the gas, a group that includes
\cdh, \cqh, and \cthd, or whether it is not, a group that includes both
complex species and hydrogenated species such as CH$_4$ and C$_3$H$_4$.  As
emphasized in Fig.~\ref{earlytimes}, the main difference is that the molecules
pertaining to the first group (\cdh\ in Fig.~\ref{earlytimes}) go through an
early phase of selective destruction before reaching steady-state, while the
molecules belonging to the second group (CH$_4$ in Fig.~\ref{earlytimes})
start with a phase of formation.

Looked at in more detail, the unsaturated hydrocarbons initially present in
the gas are quickly heavily destroyed in the very outer layers of the PDR,
while the destruction is much slower in more shielded layers.  By a time of a
few$\times 10^4$ yr, fractional abundance peaks are found at A$_V\sim1.5$--2
with a value close to that of steady-state. As time progresses, steady-state
is reached for the outer parts, while the fractional abundances in the inner
layers may drop several orders of magnitude before steady-state is reached at
$t\sim5$--$10\times10^6$ yr. The end result is a large decrease in abundance
from the initial value for the outermost layers (A$_V< 1$) and a significant
decrease, very much dependent on the molecule, for slabs with A$_V> 7$.

The molecules that were not present initially in the gas (e.g. CH$_{4}$,
CH$_2$) initially show a larger increase in the deeper layers of the cloud. We
usually find the highest abundance for each A$_V$ at a time $t\sim10^4$ yr or
shortly thereafter. After this moment, the evolution of the fractional
abundances of these molecules is similar to the one described for the previous
group: the fractional abundances do not change appreciably in the outer parts
of the PDR (A$_V\sim1$--2) while the abundances may drop, although not so
dramatically, in the inner layers until reaching steady-state conditions at
$t\sim$5--10$\times10^6$ yr. Interestingly, CH$_4$ abundance does not follow
this description completely for A$_V > 3$, where it keeps increasing until a
time $\sim5\times10^4$ yr. Then, it shows very little variation, and it only
decreases very slightly for A$_V> 8$.

\begin{table*}
 \caption[]{Peak fractional abundances and A$_V$ 
for cyanopolyynes, HCN, and HNC}
 \label{cyanopeaks}
\begin{center}
 \begin{tabular}{lcllclrcll}
\noalign{\smallskip}\hline\noalign{\smallskip}
Species & & \multicolumn{2}{c}{$t=10^4$ yr} & &
\multicolumn{2}{c}{$t=3\times10^5$ yr} & &
\multicolumn{2}{c}{$t=5\times10^6$ yr} \\
&& n(X)/n$_{\rm H}$ & A$_V$ && n(X)/n$_{\rm H}$ & A$_V$ && n(X)/n$_{\rm H}$ &
A$_V$ \\
\noalign{\smallskip}\hline\noalign{\smallskip}
HCN && $3.9\times10^{-9}$ &13.0 && $1.0\times10^{-9}$ & 8.0 && $1.5\times10^{-9}$ & 8.0\\
HNC && $3.5\times10^{-9}$ &13.0 && $9.9\times10^{-10}$ & 9.0 && $1.4\times10^{-9}$ & 8.0\\
HC$_3$N && $4.7\times10^{-11}$ &13.0 && $3.3\times10^{-13}$ & 8.0 && $2.4\times10^{-13}$ & 8.0\\
HC$_5$N && $1.1\times10^{-12}$ &13.0 && $4.5\times10^{-14}$ &13.0 && $5.4\times10^{-15}$ & 6.0\\
HC$_7$N && $4.5\times10^{-14}$ &13.0 && $2.4\times10^{-15}$ &13.0 && $5.3\times10^{-16}$ & 5.5\\
HC$_9$N && $1.7\times10^{-15}$ &13.0 && $1.8\times10^{-16}$ & 6.0 && $6.7\times10^{-17}$ & 5.5\\
\noalign{\smallskip}\hline\noalign{\smallskip}
\end{tabular}
\end{center}
\end{table*}

All hydrocarbons show a similar abundance profile at steady state (or even at
earlier times): their fractional abundances grow with depth into the cloud
until they reach a peak, after which the fractional abundance decreases with
depth more or less drastically. The peak position and the steepness of the
decrease depend on the hydrocarbon. As can be seen in Table~\ref{carbonpeaks},
the more unsaturated and simpler hydrocarbons typically peak in outer layers
of the cloud, while as the number of C atoms grows or as the molecules become
more hydrogenated, the peaks are found deeper into the cloud. For instance, by
steady state, \cdh\ peaks at A$_V=1.6$; \cthd, at A$_V=1.9$; C$_8$H, at
A$_V=2.0$; and CH$_4$, at A$_V=8$. This distribution is not continuous; there
seem to be two main regions, A$_V=1.5$--$2$ and A$_V=5.5$--$8$, where all the
peaks are found. In fact, most of the hydrocarbons show two peaks at
those two ranges, the relative importance of which seems to depend on the
number of carbon atoms present or on being or not saturated. This secondary
peak (primary in some cases) is due to the bump in C abundance shown in
Figure~1. Hydrocarbons containing fewer carbon atoms typically reach
larger fractional abundances; for example, CH$_4$, \cdh\ and \cqh\ are some of
the most abundant hydrocarbons.

\begin{table*}
 \caption[]{Peak fractional abundances and A$_V$ for other molecules typically
 detected in dense clouds}
 \label{otherpeaks}
\begin{center}
 \begin{tabular}{lclrclrclr}
\noalign{\smallskip}\hline\noalign{\smallskip}
Species & & \multicolumn{2}{c}{$t=10^4$ yr} & &
\multicolumn{2}{c}{$t=3\times10^5$ yr} & &
\multicolumn{2}{c}{$t=5\times10^6$ yr} \\
&& n(X)/n$_{\rm H}$ & A$_V$ && n(X)/n$_{\rm H}$ & A$_V$ && n(X)/n$_{\rm H}$ &
A$_V$ \\
\noalign{\smallskip}\hline\noalign{\smallskip}
NH$_3$ && $6.8\times10^{-9}$ &13.0 && $1.7\times10^{-8}$ &13.0 && $3.6\times10^{-8}$ &13.0\\
HCO$^+$ && $4.5\times10^{-9}$ &13.0 && $8.3\times10^{-9}$ &13.0 && $9.4\times10^{-9}$ &13.0\\
H$_2$CO && $1.2\times10^{-9}$ &13.0 && $1.1\times10^{-9}$ & 7.0 && $1.1\times10^{-9}$ & 7.0\\
CS && $1.5\times10^{-9}$ &13.0 && $6.3\times10^{-10}$ & 6.0 && $5.9\times10^{-10}$ & 6.0\\
N$_2$H$^+$ && $1.0\times10^{-10}$ &13.0 && $1.9\times10^{-10}$ &13.0 &&
$3.4\times10^{-10}$ &13.0\\
HCS$^+$ && $1.3\times10^{-11}$ & 2.0 && $1.5\times10^{-11}$ & 2.2 && $1.5\times10^{-11}$ & 2.2\\
\noalign{\smallskip}\hline\noalign{\smallskip}
\end{tabular}
\end{center}
\end{table*}

\subsection{HCN, HNC, and cyanopolyynes}
\label{sectcyano}

\begin{figure}
 \begin{center}
   \includegraphics[width=8.4cm]{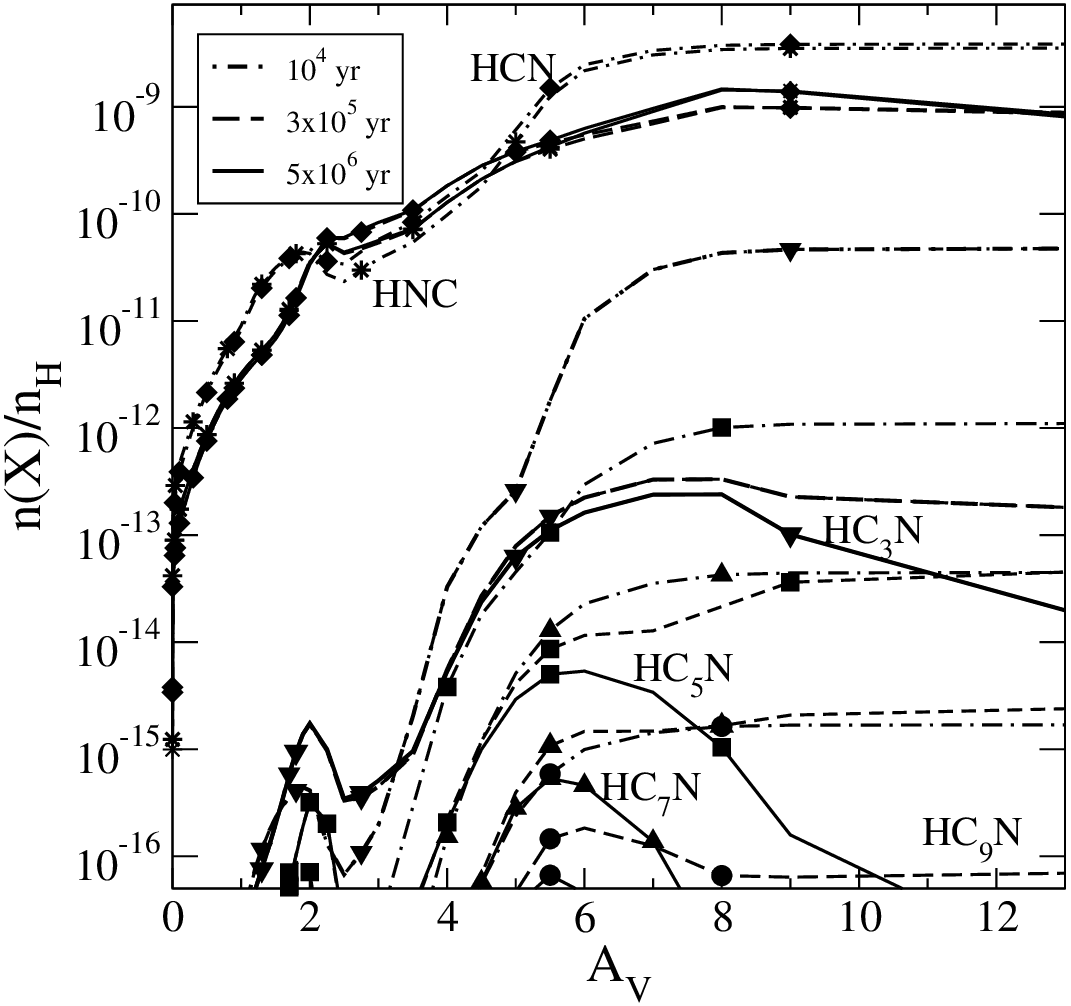}
   \caption{As Fig.~\ref{carbon1}, but for the molecules HCN (diamonds), HNC
   (stars), HC$_3$N (filled triangles pointing down), HC$_5$N (filled
   squares), HC$_7$N (filled triangles pointing up), and HC$_9$N (filled
   circles).}
   \label{cyano}
 \end{center}
\end{figure}

Fig.~\ref{cyano} shows the evolution of the fractional abundance profiles
for the HCN and HNC molecules, and the cyanopolyynes \hctn, HC$_5$N, HC$_7$N,
and HC$_9$N, while Table~\ref{cyanopeaks} lists the abundance peaks and the
A$_V$ where they are found.  Initially present in the gas, \hctn\ experiences
a progressive diminution in all the slabs as time progresses, which is
particularly severe in the first $10^5$ yr. When steady-state is reached,
\hctn\ shows a profile with two peaks, a very low one at A$_V\sim2$, and the
main one at A$_V\sim8$.  Unlike the case for hydrocarbons, the abundance of
\hctn\ in the deepest layers is less than an order of magnitude below that at
the peak.

The more complex cyanopolyynes show a behaviour similar to those unsaturated
hydrocarbons not present initially in the gas. First, they go through a phase
of formation until they achieve maximum abundances at all A$_V$ at a time
$t\sim10^4$ yr, when peak abundances are in the inner regions.  Then they are
progressively destroyed at all optical depths. The steady-state profiles of
these molecules show a low abundance peak at A$_V\sim2$, the main abundance
peak, one or two orders of magnitude over the secondary peak, at
A$_V\sim4-8$, and a decrease at deeper layers. 

The initially-present HCN and HNC molecules show a completely different but
nearly identical behaviour (see Table~\ref{cyanopeaks}). Their destruction is
very intense in the first $10^4$ yr for slabs A$_V < 2$, while only moderate
for the deeper layers. The destruction goes on until a time $t\sim10^5$ yr,
when there is a slight increase of abundances for A$_V\sim3$--8, before
steady-state is reached. An abundance peak then exists at A$_V=5.5$ and a
slightly lower, but almost uniform, abundance occurs at deeper layers. As 
happened with the cyanopolyynes, there is a small bump in abundance at
A$_V\sim2.1$, approximately 20 times lower than the main peak.

\subsection{Other molecules}
\label{sectother}

\begin{figure}
 \begin{center}
   \includegraphics[width=8.4cm]{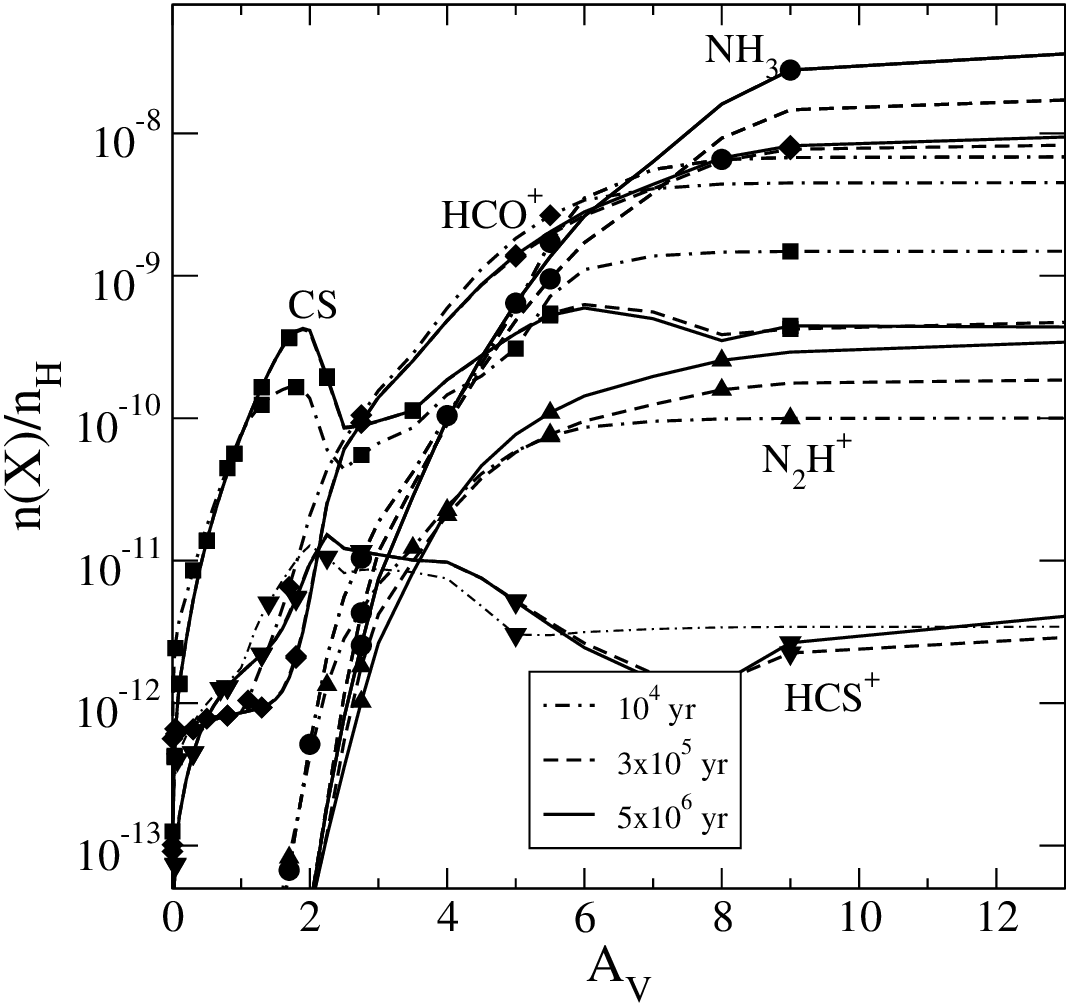}
   \caption{As Fig.~\ref{carbon1}, but for the molecules CS (filled squares),
   \hcop\ (filled diamonds), \nh\ (filled circles), \nthp\ (filled
   triangles pointing up), and \hcsp\ (filled triangles pointing down).}
   \label{other1}
 \end{center}
\end{figure}

Fig.~\ref{other1} shows the abundance profiles at assorted times of CS,
\hcop, \nh, and \nthp, all of which are well-known species in dense
regions. The peak abundances at the selected times are listed in
Table~\ref{otherpeaks}.  Present in the initial gas, CS shows a fractional
abundance profile quite similar to unsaturated hydrocarbons with two peaks of
very similar abundances, the inner one $\sim 1.5$ times larger than the
outer one, but with a very small fall in abundance for the deeper layers,
where abundance is very close to constant. 

\begin{figure}
 \begin{center}
   \includegraphics[width=\columnwidth]{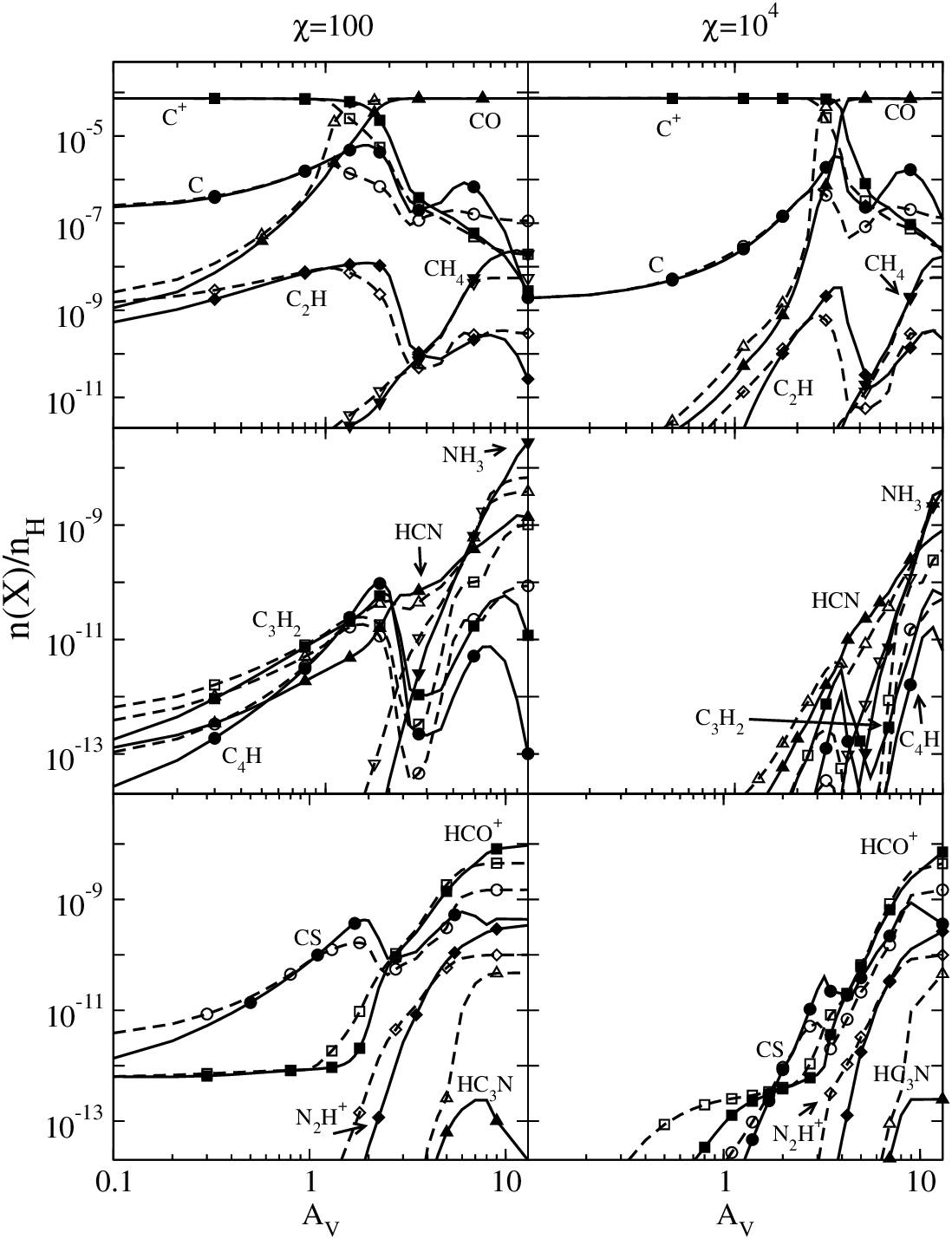}
   \caption{Comparison of the fractional abundances obtained with (left-hand
   panel) the standard model and (right-hand panel) with a model using
   $\chi=10^4$. The results are shown at an early time, $10^4$ yr (dashed
   lines and open symbols), and at $10^7$ yr (solid lines and filled symbols),
   when steady-state has been reached.}
   \label{radiation}
 \end{center}
\end{figure}

The species \hcop, \nh, and \nthp\ show profiles somewhat similar to those of
HCN and HNC.  Unlike these latter two species, however, the fractional
abundance peak of the three molecules is found in the innermost slab of our
model.

One common thread running through these descriptions of assorted profiles is
the following: the dominant features of the fractional abundance profiles as a
function of A$_V$ for most species show a strong similarity to plots of the
fractional abundance of the same molecules as a function of time in
homogeneous dark cloud models \citep*[see e.g.,][]{Taylor98}. The so-called
 `early-time' molecules, such as CS or \cdh, peak in the outer layers of the
PDR, while the  `late-time' molecules, such as \nh, \nthp, or \hcop, peak in
the innermost layers. This is not too surprising; the early-time chemistry in
the homogeneous cloud (based on a high abundance of atomic C) is similar to
that in the PDR layers with A$_V\sim$1-2, while the late-time chemistry (based
on a high abundance of CO) is similar to that in the inner layers.

\begin{table*}
 \caption{ Peak fractional abundances and A$_V$ for selected molecules for a
 model with $\chi=10^4$}
 \label{tableradiation}
 \begin{tabular}{lclrclrclr}
\noalign{\medskip}\hline\noalign{\smallskip}
Species & & \multicolumn{2}{c}{$t=10^4$ yr} & &
\multicolumn{2}{c}{$t=3\times10^5$ yr} & &
\multicolumn{2}{c}{$t=1\times10^7$ yr} \\
%\cline{3-4}\cline{6-7}\cline{9-10}
&& n(X)/n$_{\rm H}$ & A$_V$ && n(X)/n$_{\rm H}$ & A$_V$ && n(X)/n$_{\rm H}$ &
A$_V$ \\
\noalign{\medskip}\hline\noalign{\smallskip}
CH$_4$ && $5.7\times10^{-9}$ & 9.0 && $2.3\times10^{-8}$ &13.0 && $2.3\times10^{-8}$ &13.0\\
NH$_3$ && $6.7\times10^{-9}$ &13.0 && $1.1\times10^{-8}$ &13.0 && $1.9\times10^{-8}$ &13.0\\
CH && $6.0\times10^{-9}$ & 2.5 && $1.2\times10^{-8}$ & 3.0 && $1.2\times10^{-8}$ & 3.2\\
HCO$^+$ && $4.4\times10^{-9}$ &13.0 && $6.8\times10^{-9}$ &13.0 && $7.1\times10^{-9}$ &13.0\\
CH$_2$ && $4.6\times10^{-9}$ & 2.5 && $5.4\times10^{-9}$ & 3.2 && $5.5\times10^{-9}$ & 3.2\\
C$_2$H && $8.1\times10^{-10}$ & 2.5 && $3.3\times10^{-9}$ & 3.0 && $3.3\times10^{-9}$ & 3.2\\
C$_2$H$_2$ && $4.1\times10^{-9}$ &13.0 && $1.7\times10^{-9}$ &13.0 && $1.6\times10^{-9}$ &13.0\\
HCN && $3.8\times10^{-9}$ &13.0 && $1.1\times10^{-9}$ &13.0 && $1.6\times10^{-9}$ &13.0\\
HNC && $3.5\times10^{-9}$ &13.0 && $1.0\times10^{-9}$ &13.0 && $1.5\times10^{-9}$ &13.0\\
H$_2$CO && $1.2\times10^{-9}$ &13.0 && $1.1\times10^{-9}$ & 9.0 && $1.1\times10^{-9}$ & 9.0\\
CS && $1.5\times10^{-9}$ &13.0 && $9.0\times10^{-10}$ & 9.0 && $8.7\times10^{-10}$ & 9.0\\
N$_2$H$^+$ && $9.9\times10^{-11}$ &13.0 && $1.6\times10^{-10}$ &13.0 && $2.6\times10^{-10}$ &13.0\\
CH$_3$ && $9.2\times10^{-11}$ & 7.0 && $1.2\times10^{-10}$ & 6.0 && $1.2\times10^{-10}$ & 6.0\\
C$_3$H$_2$ && $1.0\times10^{-9}$ &13.0 && $7.1\times10^{-11}$ & 9.0 && $7.3\times10^{-11}$ & 9.0\\
C$_3$H && $1.9\times10^{-10}$ &13.0 && $5.5\times10^{-11}$ & 9.0 && $5.6\times10^{-11}$ & 9.0\\
C$_4$H && $8.7\times10^{-11}$ &13.0 && $1.6\times10^{-11}$ & 9.0 && $1.6\times10^{-11}$ & 9.0\\
C$_2$H$_4$ && $3.9\times10^{-13}$ &13.0 && $2.1\times10^{-12}$ &13.0 && $2.4\times10^{-12}$ &13.0\\
C$_2$H$_3$ && $1.4\times10^{-12}$ & 7.0 && $2.2\times10^{-12}$ & 8.0 && $2.2\times10^{-12}$ & 8.0\\
C$_4$H$_2$ && $2.3\times10^{-10}$ &13.0 && $2.5\times10^{-12}$ &13.0 && $1.9\times10^{-12}$ &13.0\\
C$_5$H && $1.0\times10^{-11}$ &13.0 && $8.3\times10^{-13}$ & 9.0 && $7.6\times10^{-13}$ & 9.0\\
C$_6$H && $4.1\times10^{-12}$ &13.0 && $7.9\times10^{-13}$ & 9.0 && $7.0\times10^{-13}$ & 9.0\\
HC$_3$N && $4.5\times10^{-11}$ &13.0 && $3.4\times10^{-13}$ &13.0 && $2.5\times10^{-13}$ &13.0\\
C$_3$H$_4$ && $9.8\times10^{-14}$ &13.0 && $2.3\times10^{-13}$ &13.0 && $2.0\times10^{-13}$ &13.0\\
C$_8$H && $2.0\times10^{-13}$ &13.0 && $1.3\times10^{-13}$ & 9.0 && $1.1\times10^{-13}$ & 9.0\\
C$_7$H && $4.4\times10^{-13}$ &13.0 && $1.1\times10^{-13}$ & 9.0 && $8.5\times10^{-14}$ & 9.0\\
C$_3$H$_3$ && $1.3\times10^{-14}$ &13.0 && $3.4\times10^{-14}$ & 9.0 && $4.4\times10^{-14}$ & 9.0\\
C$_5$H$_2$ && $5.1\times10^{-12}$ &13.0 && $2.9\times10^{-14}$ &13.0 && $2.8\times10^{-14}$ & 9.0\\
C$_4$H$_3$ && $1.3\times10^{-13}$ &13.0 && $2.1\times10^{-14}$ & 9.0 && $2.3\times10^{-14}$ & 9.0\\
CH$_3$C$_4$H && $1.1\times10^{-13}$ &13.0 && $1.9\times10^{-14}$ &13.0 && $1.4\times10^{-14}$ & 9.0\\
HC$_5$N && $1.1\times10^{-12}$ &13.0 && $2.6\times10^{-14}$ &13.0 && $1.3\times10^{-14}$ & 9.0\\
C$_6$H$_2$ && $5.3\times10^{-12}$ &13.0 && $1.9\times10^{-14}$ &13.0 && $1.2\times10^{-14}$ & 9.0\\
C$_6$H$_6$ && $4.7\times10^{-14}$ &13.0 && $1.1\times10^{-14}$ & 9.0 && $3.2\times10^{-15}$ & 9.0\\
\noalign{\medskip}\hline\noalign{\smallskip}
\end{tabular}
\end{table*}

\section{Influence of physical parameters}
\label{sct_disc}

\subsection{Radiation field}

In order to determine how the external radiation field affects the abundances
of the different species, especially the hydrocarbons, we ran three additional
models, each with a different intensity of the radiation field, indicated by
scaling factors of $\chi=10, 10^3$, and $10^4$. Fig.~\ref{radiation}
compares the results of our standard model with the results obtained with a
model using $\chi=10^4$ at an early stage of evolution, $10^4$ yr, and at
$10^7$ yr, when steady-state has been reached for all
A$_V$. Table~\ref{tableradiation} lists the peak abundances of a variety of
species for three times when $\chi=10^4$.

The evolution of the fractional abundances when $\chi = 10^4$ differs most
strongly from the standard case, as the right-hand side of
Fig.~\ref{radiation} shows, in the sweeping of most of the molecular species
from the outer layers by the intense radiation.  Most of the molecules shown
have already disappeared from the layers with $A_V < 1$ by $t=10^4$ yr and,
when steady-state is reached at $10^7$ yr, there are no significant molecular
abundances at slabs with A$_V<1$. Even molecular hydrogen disappears from the
external layers as the position at which H$_2$ becomes the most abundant form
of hydrogen increases to A$_V\sim1.3$ when $\chi=10^4$. Thus, for the model
with $\chi=10^4$, the gas at A$_V<1$ is mostly in atomic form. The fractional
abundance peaks of the species with profiles that peak in the outer layers are
consequently found deeper into the cloud at all times, the typical effect
being an increase from A$_V\sim1.5$--2 to A$_V\sim3$--4 (and the inner maxima
and/or enhancements from A$_V\sim4.5$--7 to A$_V\sim7$--9).  On the other hand,
the results for the model with a less intense radiation field than the
standard model, $\chi=10$, are the opposite and many molecular species have
significant fractional abundances at A$_V$ as low as 0.1. The fractional
abundance peaks are then located in more external layers of the cloud.

The maximum fractional abundances are also modified by the intensity of the
radiation field in all of the cases. As $\chi$ is increased from 100 to
10$^4$, the outer peak abundance of hydrocarbons is typically reduced by a
factor of $\la10$, but the decrease can range from very little, as for CH$_2$,
to two (\cqh\ or C$_6$H) or three (C$_8$H) orders of magnitude.  On the other
hand, the  `inner' abundance peak of hydrocarbons increases by a small
factor, between 2 and 10, and in some cases, such as \cthd, it can become
larger than the (outer) peak abundance in the standard model.

The peak fractional abundances of the cyanopolyynes have a much greater
dependence on the radiation field, and we find general enhancements of two and
three orders of magnitude for the outer layers, when $\chi=10^4$. This
increase does not change the fact that there are only traces of these
molecules in these layers. The peak abundance at high visual extinction also
increases with a larger radiation field, but only by small factors (2--3).

The radiation field is able to modify the chemistry even in deep slabs of the
cloud.  We find that even if the changes in intermediate layers
(A$_V\sim5$--8) can be small, a factor of a few, they are non-zero; only the
gas at A$_V \ga 9$--10 is completely shielded from the radiation for all the
models and does not show any significant changes at any time between the
different models. As a result, for the two radiation fields considered, all
the molecules that reach their maximum abundances in the inner cloud, such as
HCN, \nh, and \nthp, have approximately the same abundances at high visual
extinction at steady-state and at intermediate times. On the other hand, these
same molecules, which were not abundant in the outer layers of the cloud for
$\chi=100$, completely disappear from the superficial layers for $\chi=10^4$.

\subsection{Density and temperature}

In order to explore the effects of different densities and temperatures, we
reran our standard model, modifying either the density or the temperature of
the gas. The densities and temperatures used were $n_{\rm H} =10^3, 10^5$, and
$10^6$\cmt, and $T= 10$, 100, and 200 K.

The effects of density change bear some similarities with the effects of the
radiation field: a higher density increases the shielding and the abundances of
most of the molecules in the outer layers of the cloud; a lower density on the
other hand, clears these same layers of much molecular material, and displaces
the peak abundance position to deeper parts of the cloud. The results are not
identical, particularly in the way the inner and outer peaks of many molecules
change with density.  

Unsaturated hydrocarbons (\cdh, \cqh, C$_6$H, C$_8$H) show an increase in the
peak abundance in the outer abundance peak with increasing density, that can
range from a few percent (\cdh) to an order of magnitude (\cqh, C$_6$H); and a
simultaneous decrease in the abundance of the inner abundance peak, which can
be almost three orders of magnitude for the molecules containing more than a
few carbon atoms. The resulting effect is that at high densities there is only
a narrow range of A$_V$ in the outer layers of the cloud where these molecules
show a significant abundance; the more carbon atoms they contain, the narrower
the range. At the lowest density, on the other hand, the inner abundance peak
may not only be larger than the outer abundance peak, but also larger than the
peak abundance in the standard model. Hydrocarbons that peak inside the
cloud, such as methane and C$_3$H$_4$, also show a clear increase in abundance
as density increases. At $n=10^6$\cmt, methane also shows an outer peak.

The rest of the molecules we have discussed in Section~\ref{sct_results} also
show clear trends in peak abundance with a change in density. The
cyanopolyynes show a steep increase in the abundance in the outer layers as
density increases (it can reach three or more orders of magnitude), but never
exceed the inner peak abundance, which does not change much and is found at
slightly lower A$_V$. At the higher density, \hctn\ reaches 10$^{-12}$\cmt\
for times $\sim 5\times10^5$ yr. The abundance of the radical CS increases
slightly with density (up to a factor of 4 in the outer parts of the cloud),
whereas \hcop\ shows a behaviour similar to the unsaturated hydrocarbons, but
with changes typically less than an order of magnitude: the peak abundance at
high density is a factor $\sim8$ lower than for the standard model. \nh\ and
\nthp\ show similar trends as \hcop, but while the fractional abundance
decrease at $n=10^6$\cmt\ is less than a factor of 3 for \nh, it is larger
than 30 for \nthp.

\begin{table}
\caption{Atomic initial abundances}
\label{table_elemental}
\begin{tabular}{ll}
Element & n(X)/n$_{\rm H}$ \\
\noalign{\smallskip}\hline\noalign{\smallskip}
H     & 1\\
He    & \hspace{0mm}0.14 \\
C$^+$ & $7.3\times10^{-5}$\\
N     & $2.14\times10^{-5}$\\
O     & $1.76\times10^{-4}$\\
S$^+$ & $8\times10^{-8}$\\ 
Si$^+$ & $8\times10^{-9}$\\
Fe$^+$ & $3\times10^{-9}$\\ 
Na$^+$ & $2\times10^{-9}$\\
Mg$^+$ & $7\times10^{-9}$\\ 
P$^+$ & $3\times10^{-9}$\\
Cl$^+$ & $4\times10^{-9}$\\
\noalign{\smallskip}\hline\noalign{\smallskip}
\end{tabular}
\end{table} 

Variations in the peak fractional abundance with temperature as we raise $T$
to 200 K tend to be less dramatic than with density. The differences in
fractional abundance in the outer layers relative to our standard model are
rarely larger than a factor of 2, and often show less than a 50 per cent
change. The differences are larger for the inner layers, where a few molecules
show a change of more than an order of magnitude for $T=200$ K. Most of the
carbon chains (except \cdh\ or C$_6$H) show a large decrease of abundance
($\sim100$) in the innermost layers as $T$ increases, whereas for the outer
layers only \cdh, C$_6$H, and C$_8$H show an increase in abundance with
temperature. The fractional abundances of CH$_4$, H$_2$CO, CS, and the
cyanopolyynes \hctn\ and HC$_5$N also diminish with increasing $T$, while the
abundances of HCN and HNC increase in the outer layers with $T$ (a factor of
$\sim3$ at $T= 200$~K), but decrease in the inner layers (less than 40 per
cent). The ions \hcop\ and \nthp\ show increasing abundance with increasing
temperature (a factor of $\sim 2$ for $T=200$~K).  The abundances of the rest
of the molecules that we discussed in Section~\ref{sct_results} show small
changes of a few percent.

\subsection{Initial abundances}
\label{sectelemental}

The previous models were all run assuming as initial abundances those expected
if a dense PDR is created by a star recently formed inside a molecular
cloud. In order to explore the dependence of the model on the initial
fractional abundances, it is also interesting to determine what are the
differences between our standard model and other models that have
different initial fractional abundances.

\begin{figure}
 \begin{center}
   \includegraphics[width=8.4cm]{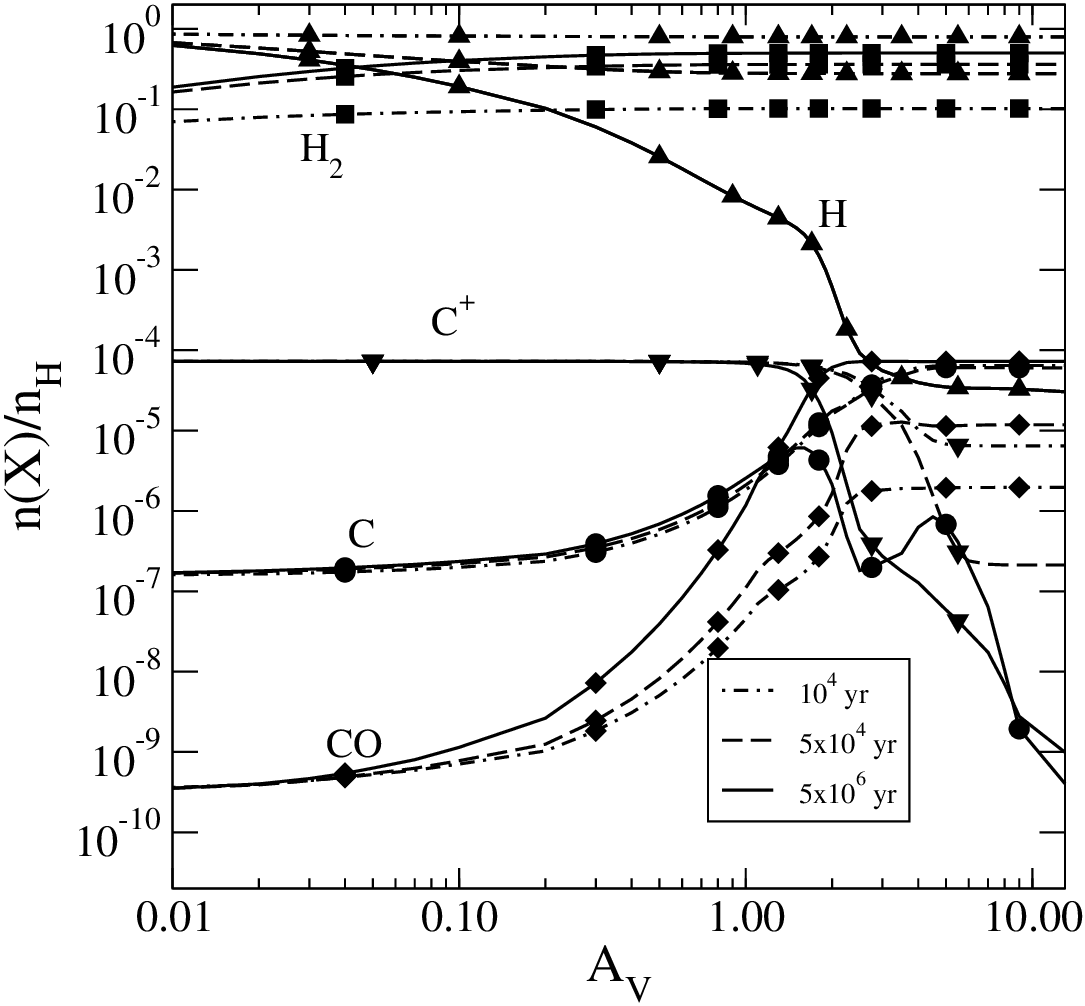}
   \caption{As Fig.~\ref{bigones} for a model with atomic initial abundances.}
   \label{bigonesatomic}
 \end{center}
\end{figure}

\begin{figure}
 \begin{center}
   \includegraphics[width=8cm]{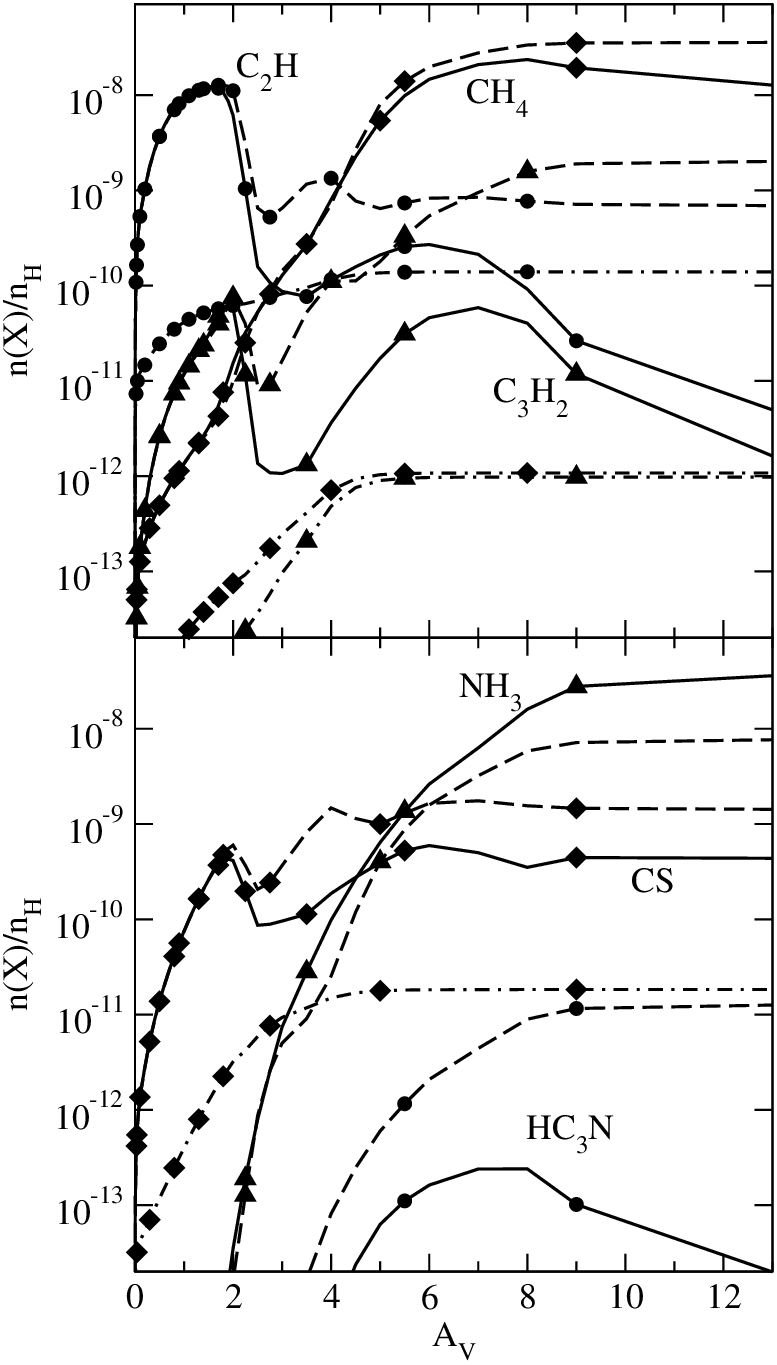}
   \caption{Fractional abundance profiles for a model with atomic initial
   abundances and $\chi=100$ at three times: $10^4$ yr (dot-dashed lines),
   $3\times10^5$ yr (dashed-lines) and $5\times10^6$ yr (solid lines). Some
   hydrocarbons are shown on the top figure; viz., \cdh\ (filled
   circles), \cthd\ (filled triangles up), and CH$_4$ (filled diamonds). On
   the bottom figure are depicted \hctn\ (filled circles), \nh\ (filled
   triangles up), and CS (filled diamonds).}
   \label{elemental}
 \end{center}
\end{figure}

\begin{table*}
 \caption{Peak fractional abundances and A$_V$ for selected molecules for a
 model with  atomic initial abundances and $\chi=100$}
\label{atomicpeaks}
 \begin{tabular}{lclrclrclr}
\noalign{\medskip}\hline\noalign{\smallskip}
Species & & \multicolumn{2}{c}{$t=10^4$ yr} & &
\multicolumn{2}{c}{$t=3\times10^5$ yr} & &
\multicolumn{2}{c}{$t=5\times10^6$ yr} \\
&& n(X)/n$_{\rm H}$ & A$_V$ && n(X)/n$_{\rm H}$ & A$_V$ && n(X)/n$_{\rm H}$ &
A$_V$ \\
\noalign{\medskip}\hline\noalign{\smallskip}
NH$_3$ && $4.0\times10^{-15}$ &13.0 && $7.7\times10^{-9}$ &13.0 && $3.6\times10^{-8}$ &13.0\\
CH$_4$ && $1.1\times10^{-12}$ &13.0 && $3.6\times10^{-8}$ &13.0 && $2.4\times10^{-8}$ & 8.0\\
CH && $1.8\times10^{-10}$ & 1.1 && $1.9\times10^{-8}$ & 1.8 && $1.9\times10^{-8}$ & 1.7\\
C$_2$H && $1.4\times10^{-10}$ &13.0 && $1.3\times10^{-8}$ & 1.8 && $1.2\times10^{-8}$ & 1.6\\
HCO$^+$ && $7.3\times10^{-12}$ &13.0 && $6.7\times10^{-9}$ &13.0 && $9.4\times10^{-9}$ &13.0\\
CH$_2$ && $3.7\times10^{-12}$ & 1.0 && $6.4\times10^{-9}$ & 1.7 && $7.3\times10^{-9}$ & 1.8\\
C$_2$H$_2$ && $5.2\times10^{-12}$ &13.0 && $1.1\times10^{-8}$ &13.0 && $1.5\times10^{-9}$ & 8.0\\
HCN && $9.2\times10^{-13}$ &13.0 && $9.6\times10^{-10}$ &13.0 && $1.5\times10^{-9}$ & 8.0\\
HNC && $7.4\times10^{-13}$ &13.0 && $9.0\times10^{-10}$ &13.0 && $1.4\times10^{-9}$ & 8.0\\
H$_2$CO && $1.5\times10^{-12}$ &13.0 && $1.2\times10^{-9}$ & 7.0 && $1.1\times10^{-9}$ & 7.0\\
CS && $1.8\times10^{-11}$ &13.0 && $1.8\times10^{-9}$ & 7.0 && $5.9\times10^{-10}$ & 6.0\\
N$_2$H$^+$ && $1.9\times10^{-15}$ & 7.0 && $1.2\times10^{-10}$ &13.0 && $3.4\times10^{-10}$ &13.0\\
CH$_3$ && $4.9\times10^{-12}$ & 7.0 && $2.7\times10^{-10}$ & 3.5 && $2.4\times10^{-10}$ & 2.2\\
C$_4$H && $3.7\times10^{-11}$ &13.0 && $5.0\times10^{-10}$ & 7.0 && $9.6\times10^{-11}$ & 1.8\\
C$_3$H && $1.2\times10^{-12}$ &13.0 && $5.7\times10^{-10}$ & 9.0 && $8.7\times10^{-11}$ & 2.0\\
C$_3$H$_2$ && $9.8\times10^{-13}$ &13.0 && $2.0\times10^{-9}$ &13.0 && $6.3\times10^{-11}$ & 1.9\\
C$_5$H && $1.6\times10^{-12}$ &13.0 && $2.7\times10^{-10}$ & 6.0 && $9.9\times10^{-12}$ & 2.0\\
C$_2$H$_3$ && $3.8\times10^{-17}$ & 0.9 && $7.0\times10^{-12}$ & 2.2 && $8.0\times10^{-12}$ & 2.0\\
C$_6$H && $3.3\times10^{-13}$ &13.0 && $3.4\times10^{-10}$ & 6.0 && $3.2\times10^{-12}$ & 1.9\\
C$_2$H$_4$ && $9.7\times10^{-20}$ &13.0 && $5.4\times10^{-12}$ &13.0 && $1.9\times10^{-12}$ & 8.0\\
C$_4$H$_2$ && $6.9\times10^{-12}$ &13.0 && $1.4\times10^{-9}$ &13.0 && $1.8\times10^{-12}$ & 7.0\\
C$_8$H && $6.9\times10^{-15}$ &13.0 && $1.5\times10^{-10}$ & 5.0 && $2.7\times10^{-13}$ & 2.0\\
HC$_3$N && $1.1\times10^{-17}$ &13.0 && $1.3\times10^{-11}$ &13.0 && $2.4\times10^{-13}$ & 8.0\\
C$_3$H$_4$ && $9.2\times10^{-21}$ &13.0 && $6.4\times10^{-12}$ &13.0 && $1.9\times10^{-13}$ & 8.0\\
HC$_5$N && $6.0\times10^{-17}$ &13.0 && $3.0\times10^{-11}$ &13.0 && $5.4\times10^{-15}$ & 6.0\\
C$_6$H$_6$ && $5.0\times10^{-20}$ &13.0 && $3.6\times10^{-12}$ &13.0 && $2.1\times10^{-15}$ & 7.0\\
\noalign{\medskip}\hline\noalign{\smallskip}
\end{tabular}
\end{table*}

First, we compare our standard model with a model starting from gas that is
atomic in nature.  Figures~\ref{bigonesatomic} and \ref{elemental} show some
results obtained with a model identical to our standard model, but using the
initial abundances listed in Table~\ref{table_elemental}, derived from `low
metal' abundances \citep{Ruffle00}. The major difference between
Fig.~\ref{bigonesatomic} and its analogous, Fig.~\ref{bigones}, is that, with
standard initial abundances, H$_2$ decreases at low A$_V$ as a function of
time whereas, with atomic initial abundances, H$_2$ must be synthesized and
increases at all A$_V$ as a function of time.  Fig.~\ref{elemental} clearly
shows that the different initial abundances determine the early evolution of
the gas. The first years are obviously spent in the process of building more
complex molecules from the available elements, while, in our standard model, a
general destruction of initially present molecules is predominant for very
early times. Table~\ref{atomicpeaks} lists the maximum abundances and peak
positions of several interesting species at selected times. Much lower than in
the standard model for a time $t=10^4$ yr, the fractional abundances show an
increase as A$_V$ grows until a plateau is reached. By 1-3$\times10^5$ yr, the
fractional abundances of some molecules, such as many carbon chains and
cyanopolyynes, can be much larger at A$_V>3$ than for the standard model at
the same times, from factors of 2-3 to more than $\sim1000$. Eventually, after
a few $\times10^5$ yr when most of the C begins to be locked in the CO
molecule, the abundances for many molecules in the inner layers follow the
same evolutionary path as in the standard model and start decreasing, while
there is still a slight increase in the molecular abundances in the outer
layers. The net effect is a change from the `plateau' profile to the same
peaked profiles we observe for most species in our standard
model. Interestingly, the abundances of \hcop, \nh, and \nthp, which peak in
the innermost layers, are smaller than the ones in the standard model for
times $t<1-2\times10^6$ yrs. Steady state is reached at $t\sim10^7$ yr at all
A$_V$. The abundances at steady state are the same or almost identical ($<1$
per cent) to the fractional abundances of the standard model.

\begin{figure}
 \begin{center}
   \includegraphics[width=7cm]{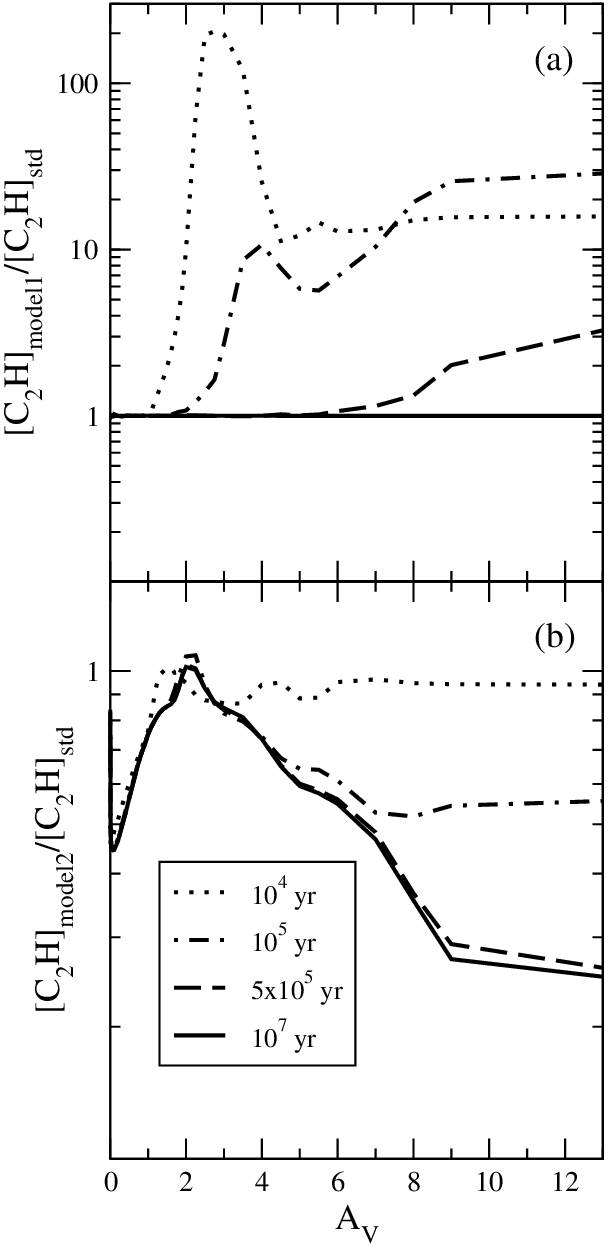}
   \caption{(a) (upper panel) \cdh\ abundance in Model CO-1 compared with our
     standard model as a function of A$_V$ at four times: $10^4$ yr (dotted
     lines), $10^5$~yr (dot-dashed lines), $5\times10^5$ yr (dashed-lines) and
     $10^7$ yr (solid lines); (b) (lower panel) as in a) but for Model CO-2.}
   \label{figratios}
 \end{center}
\end{figure}

As we pointed out in Section~\ref{sct_model}, the CO initial fractional
abundance of our standard model is greater than the one measured in TMC-1P by
a factor $\sim2$.  In order to see how this can affect our conclusions, we ran
two additional models that used the CO abundance listed by \citet{Smith04} for
TMC-1: i) a model (Model CO-1) maintaining the total standard
elemental amount of C and O, and ii) a model (Model CO-2) not
maintaining either the C/H ratio of $7.3\times10^{-5}$ or the O/H ratio of
$=1.76\times10^{-4}$. Table~\ref{tabcmodels} shows the initial abundances of
C, O, and CO for these models, as well as the overall C/H, O/H, and C/O
elemental abundance ratios, designated with the subscript  `\textit{elem}'.
As an example, Fig.~\ref{figratios} shows the \cdh\ abundances in these two
models relative to the \cdh\ abundance in our standard model.

For the first model, the more abundant initial free C in the gas produces a
smaller abundance of CO for times $\la 2\times10^5$ yrs and a correspondingly
larger abundance of C and C$^+$ up to 3--5$\times10^5$ yrs. This effect is
particularly important for atomic C in the inner layers. For early times, the
free carbon produces an increase of the abundance of all the carbon-chain
molecules, at all layers $A_V\ga2$. These differences in abundance relative to
the standard model diminish with time from the outer to the inner layers, and
last up to 5--10$\times10^5$ yrs for layers A$_V \ga 3-4$. The differences in
abundance at the positions of the external peak abundances of the standard
model are generally very small. For early times, the abundances at layers with
$2\la$A$_V \la 5$ can be significantly larger by factors between a few (for
CH$_4$) and up to $10^3-10^4$ (for C$_6$H), while for layers with $A_V\ga5$,
most of the carbon-chain molecules show abundances that range from $\sim25$ to
a few hundred times larger than the ones in the standard model at early
times. Other carbon-containing molecules show a similar behaviour. On the
other hand, HCO$^+$ and \nh\ show smaller abundances, by factors of 3--4, for
layers A$_V<6$. For times later than $\sim10^6$ yr, the abundances are
practically identical to the ones from the standard model, as we expected from
the results of the model with initial atomic abundances.

\begin{table}
\caption{Initial abundances of C, O, and CO [$n(X)/n_{\rm H}$] for various
    models}
\label{tabcmodels}
\begin{tabular}{lccc}
\noalign{\smallskip}\hline\noalign{\smallskip}
species & Standard Model & Model CO-1 & Model CO-2\\
\noalign{\smallskip}\hline\noalign{\smallskip}
 CO & $7.27\times10^{-5}$ & $4.00\times10^{-5}$ & $4.00\times10^{-5}$ \\
 C  & $2.81\times10^{-8}$ & $3.27\times10^{-5}$ & $2.81\times10^{-8}$ \\
 O  & $1.03\times10^{-4}$ & $1.36\times10^{-4}$ & $1.03\times10^{-4}$ \\
(C/H)$_{\rm elem}$ & $7.27\times10^{-5}$ & $7.27\times10^{-5}$ &
$4.00\times10^{-5}$ \\ 
(O/H)$_{\rm elem}$ & $1.76\times10^{-4}$ & $1.76\times10^{-4}$ &
$1.43\times10^{-4}$ \\
(C/O)$_{\rm elem}$ & 0.41 & 0.41 & 0.28 \\
\noalign{\smallskip}\hline\noalign{\smallskip}
\end{tabular}
\end{table}

Although the late-time abundances for the second model are relatively close,
generally less than an order of magnitude, to those of our standard model, the
early- and late-time behaviour of most of the species is opposite to that of
the first model, i.e. an even closer agreement to our standard model is found
at early times.  At later times, the lower general abundance of elemental C
produces smaller abundances compared to the standard model for all the
carbon-chain molecules. This effect is specially apparent in the inner layers,
A$_v>3$, with factors that range from 2 to more than 10. Also at later times,
the differences in the outer layers are usually less than 50 per cent, and in
some cases we can see a slight increase in abundance, as is the case for
\cthd\ or C$_6$H, but never affecting the abundance peaks of the standard
model. This slight increase can also be seen for HNC, HC$_3$N, and \hcop. The
abundance of \nh\ is larger by factors of 1.5--4 at all layers.

\section{Comparison with observations of the Horsehead nebula}
\label{hhnebula}

One of the main difficulties in the modelling of PDRs is to compare the
results with actual observations, because it is usually extremely difficult to
determine the distribution of the gas density, and the way the far-UV
radiation penetrates the cloud. In recent times, the Horsehead nebula has been
used as a benchmark for PDR models, because it appears to be viewed
edge-on and thus offers a good opportunity to study the structure of a PDR.

The Horsehead nebula (B33) is located at the western edge of the molecular
cloud L1630, at the near side of the H\,{\small II} region IC 434. The
Horsehead is thought to be an early Bok globule; i.e., a higher density cloud
core which is emerging from its parental and more tenuous cloud as the UV
radiation from the O9.5V star $\sigma$ Ori further to the west erodes the
surrounding gas \citep{Reipurth84}. The Horsehead nebula has been extensively
studied at many different wavelengths, from the early maps in CS, CO, and
[C\,{\small II}] (\citealt*{Lada91}; \citealt{Zhou93}; \citealt*{Kramer96}) to
more recent observations ranging from the mid-IR to the millimetre
(\citealt{Abergel02,Abergel03}; \citealt*{Pound03}; \citealt{Teyssier04};
\citealt{Habart05}; \citealt{Pety05}). The gas density derived from these
observations, either from the excitation of molecular lines or from the
penetration depth of the UV radiation, is 1--$4\times10^4$\cmt\
\citep{Zhou93,Kramer96,Abergel03}. \citet{Habart05} found that the gas density
follows a steep gradient at the cloud edge, rising to $n_{\rm H} = 10^5$\cmt\
in less than 0.02 pc, and at the same time the gas temperature changes from
$\sim200$--300 K to $\sim10$--20 K. \citet{Abergel03} derived a kinetic
temperature from single-dish observations of 30--40 K, which would correspond
to the bulk of the cloud, while the warm UV-illuminated edge would have higher
temperatures \citep{Pety05}. The far-UV incident radiation field can be
considered to have a mild intensity, $\chi \sim60$ in Draine units, and has
also been estimated in the $\chi \sim 20$--200 range
\citep{Zhou93,Kramer96,Abergel03}.

\citet{Teyssier04} and \citet{Pety05} studied the physical and chemical
structure of B33 at millimetre wavelengths, using moderate-angular-resolution
single-dish and high-angular-resolution interferometric observations,
respectively. \citet{Teyssier04} observed the emission of a variety of
molecules (C$^{18}$O, \cdh, C$_3$H, \cqh, \cthd, C$_3$H$_4$, C$_6$H,
HN$^{13}$C, \hctn, and CS) at three positions in the Horsehead nebula, one
coinciding with the peak in IR emission (IR peak), another, very close by, at
the position at the hydrocarbon peak emission, and the last one at the
position of the CO peak. These and previous observations showed a relative
stratification of H$_2$ and small hydrocarbons: the boundary of the H$_2$
emission corresponds to the edge of the PDR, while the hydrocarbon peaks,
which are well correlated with one another, are located farther into the
cloud. In addition, the CO peak occurs at a place tangential to the
hydrocarbon peak.  \citet{Teyssier04} used steady-state PDR models with the
same physical conditions as ours to explain the original observations, with
reaction rates from two different networks, UMIST95 \citep*{Millar97} and
osu.2003 \citep{Smith04}. They reproduced some of the observed fractional
abundances (C$^{18}$O and \cdh) and some column density ratios, but their
calculated peak fractional abundances for c-\cthd\ and \cqh\ are low by an
order of magnitude. The discrepancies were attributed to the return of small
carbon molecules to the gas phase through the sputtering of small carbon
grains and PAHs.

The high-angular-resolution observations of \citet{Pety05} determined the
column density and fractional abundance of C$^{18}$O, \cdh, \cqh, and c-\cthd\
at three positions, named `IR edge', `IR peak', and `cloud', in order of their
depth inside the cloud.  The `IR-edge' position corresponds to the edge of the
PDR, where the UV radiation is most intense; the `IR peak' position is located
5 arcsec east, where both the IR and hydrocarbon emission peaks of
\citet{Teyssier04} were found; and the `cloud' position is $\sim15$ arcsec
deeper into the cloud. The gas temperature estimates for the three positions
are 100, 60, and 40 K. \cite{Pety05} also used different models to try to fit
the more spatially detailed observations but the main results are not
significantly different from the ones of \citet{Teyssier04}.  Recently,
\citet{Pety07} published more data for CS, \hcsp, and \hcop\ at three
positions, one approximately at the `cloud' position and two deeper into the
molecular cloud.

\begin{figure}
  \includegraphics[width=8.4cm]{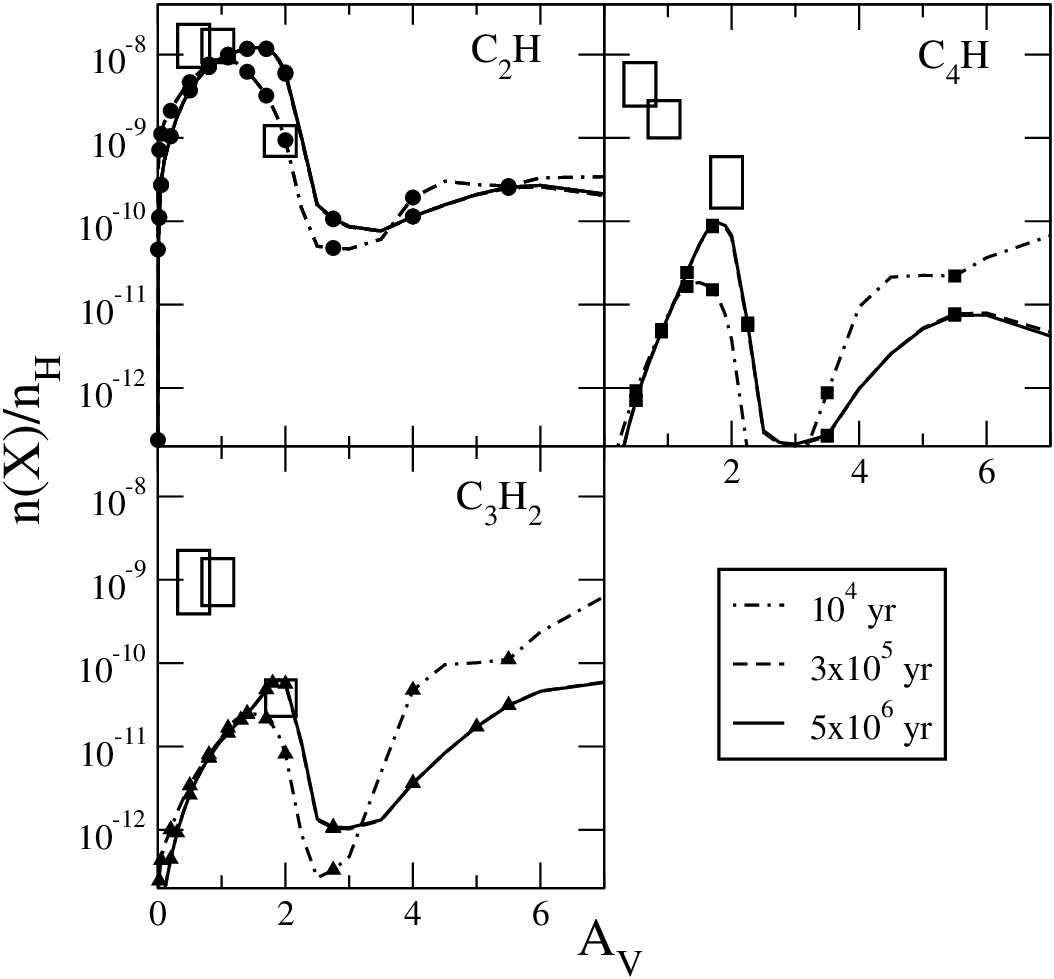}
  \caption{Comparison of the abundance profiles of our standard model for
  \cdh, \cqh, and \cthd\ at three times with the values observed by
  \citet{Pety05} at the three positions, labelled from left to right as
   `IR-edge',  `IR-peak', and  `cloud'. The boxes indicate the
  uncertainties in the observed fractional abundances and in the position
  (beamwidth).}
  \label{comparison1}
\end{figure}

\begin{figure}
  \includegraphics[width=8.4cm]{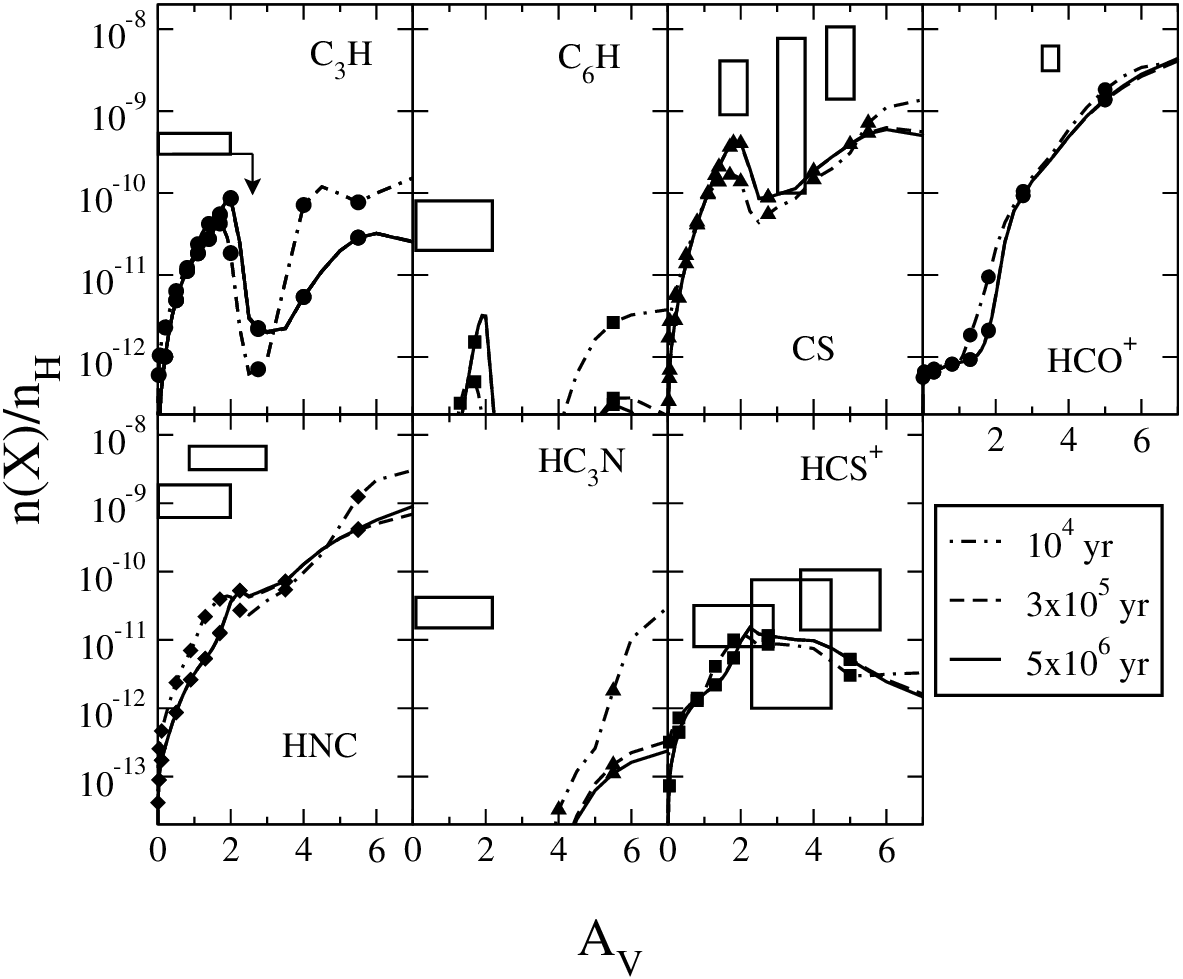}
  \caption{Similar to Fig.~\ref{comparison1}, but for C$_3$H, C$_6$H, HNC,
  and HC$_3$N in the  `IR peak' or  `hydrocarbon peak' positions observed by
  \citet{Teyssier04} and for CS, \hcop, and \hcsp\ taken from \citet{Pety07}.}
  \label{comparison2}
\end{figure}

The results of our standard model when steady-state is reached are generally
similar to those of \citet{Teyssier04} and \citet{Pety05} (with either
reaction network) concerning the details of the PDR zones and the positions
and heights of the hydrocarbon peaks.  We compare in Figs~\ref{comparison1}
and \ref{comparison2} the abundance profiles of our standard model at three
different times with the fractional abundances of some molecules observed by
\citet{Pety05} and \citet{Teyssier04}, taking into account the errors
associated with the observations.  In these figures, fractional abundance is
plotted against visual extinction, determined from position for the
observational values by assuming a constant density of $2 \times 10^{4}$
cm$^{-3}$.  The peaking of fractional abundances towards the outer layers of
the PDR in both observation and theory is striking.

Let us first consider the results at steady-state ($5 \times 10^{6}$ yr),
focusing initially on the molecules studied at high resolution by
\cite{Pety05}. The peak fractional abundances of \cdh, \cqh, and \cthd\ in our
standard model are similar to those of the model of \citet{Teyssier04}.
Compared with observation (see Fig.~\ref{comparison1}), we find reasonably
good agreement at all three observational positions for \cdh\ (less so for the
 `cloud' position, although a slight displacement of $\Delta$A$_V \sim 0.25$
would fit the three points at the same time). For \cqh\ and \cthd, the value
observed at the  `cloud' position is the only one matched by our model,
within a factor of 2 for \cqh. The underproduction for the other two positions
is $\sim 10^2-10^3$ times for \cqh\ and $\la 50$ times for \cthd.

Fig.~\ref{comparison2} show the additional species detected by
\citet{Teyssier04} and \citet{Pety07} compared to the results of our standard
model. We note that their so-called IR peak and hydrocarbon peak positions are
assumed to both be represented by visual extinctions from 0 to 2.  To convert
the actual observation of HN$^{13}$C and H$^{13}$CO$^+$ into the normal
isotopologue, we assumed a $^{12}$C/$^{13}$C ratio of 62 \citep{Langer93}.  As
can be seen, the steady-state results of our standard model range from good to
poor. CS and \hcsp\ are reproduced within a factor of $\sim3$, which for the
first could probably be attributed to the effects of sulfur depletion.  C$_3$H
and C$_6$H are underproduced in the model, similarly to the other carbon
chains, by factors of $\sim3$ and $\sim 6$, respectively.  The abundances of
\hcop\ and HNC are almost a factor of 10 below the observed value, although
there are values in the model at high A$_V$ that match the observations. We
find the largest differences for HC$_3$N, which is heavily underproduced in
the standard model at steady state.

If we consider earlier times, we find that for times of 10$^{4}$ yr and later,
most of the time dependence occurs in inner portions of the PDR, such as the
 `cloud' position of \citet{Pety05} and even larger values of visual
extinction.  For the `cloud' position, we find that the abundance profiles
of \cdh\ may be somewhat better fitted at earlier times.  In general, the
results of \citet{Teyssier04} do not extend far enough into the cloud to be
sensitive to time. There is a large time dependence for the abundance of
\hctn\ but this occurs at high values of visual extinction for the most
part. Since variations in the CO initial abundances discussed in
Sect.~\ref{sectelemental} mainly affect abundances in inner layers, with
little or no impact on the outer peak abundances, these variations do not
greatly affect the level of agreement between model and observation.

\subsection{Other physical parameters}
\label{sect_hh_td}

We used the alternative models discussed in Sect.~\ref{sct_disc} to determine
if there is a set of physical parameters that agrees better than the standard
model with the abundances derived from the observations in the Horsehead
nebula. We also ran additional models that looked at intermediate values of
the parameters or that changed initial conditions.  We did not find a unique
best-fit model.  Since different authors \citep[e.g.,][]{Pety05,Goicoechea07}
have proposed different elemental abundances to try to explain the abundances
in PDRs, we have also changed the relative elemental abundance, mainly by
increasing the abundance of C, but also of O, S, and N. An initial carbon
abundance twice of that in our standard model shows some of the better
agreements for \cqh\ (for the innermost position), C$_6$H, and especially CS
and \hcsp.   `Carbon-rich' abundances similar to the ones used by
\citet{Millar00} overproduce the abundances of most of the carbon-containing
molecules. Different sulfur abundances can be used to fine tune the amounts of
CS and \hcsp.  We also explored the effects of different values of the cosmic
ray ionization rates. We used values of 0.5, 10, 50, and 100 times the
standard value of $\zeta$. Typically, higher values of the cosmic-ray
photodissociation rate ($\ge10-50$) show better agreements for many of the
observed molecules, especially combined with a slightly higher density and
temperature and/or slightly less intense radiation field.

\subsection{Age of the PDR}

The time dependence of our models, as is also the case for the time-dependent
model of UCL \citep{Bell05}, can in principle be used to try to set an age for
a dense PDR, given a favourable geometry.  For the Horsehead nebula,
\citet{Pound03} argue that, from its size and velocity gradient, the formation
time scale could be as short as a few $\times10^5$ yr, and the object is
likely to be destroyed in $5\times10^6$ yr. These time scales are shorter or
comparable to the time needed by our standard model to reach
steady-state. But, in trying to distinguish between steady-state and earlier
times by chemical abundance profiles, we must keep in mind some formidable
obstacles. First, the uncertainties in the determination of the PDR geometry
and density structure make such an analysis very difficult because they
obscure purely temporal effects. Secondly, there is a dependence of early-time
results on what is chosen for initial abundances.  Thirdly, the abundance
profiles of the many hydrocarbons that peak in the outer layers of the cloud
show greater sensitivity to time in the inner layers of the PDR because
steady-state takes more than 10 times longer to reach in the inner
slabs. Unless the PDR is very young, these inner slab abundances are likely to
be much smaller than peak values and thus hard to observe. Moreover, most of
the more abundant molecules that peak in the deep slabs of the cloud also
reach values close to those of steady-state in a relatively short time and may
not help much in the determination of age. Some molecules, such as \hctn,
suffer large variations in fractional abundance in the same span of time and
in the position of the peak abundance, and these could be useful to carry out
age estimates. In some of the models we ran, we find better agreements for
some molecules at times $0.4-3\times10^5$ yrs, but given the uncertainties in
the observations and in the density structure of the cloud, we need more
observations of different positions and molecular transitions to have a
clearer view of the structure of the PDR and to be able to make a more precise
estimate of its age. In particular, we would need observations of the layers
with A$_V\ga4$, where we find, in our model, the largest variations of
abundance for most of the observed molecules.

\section{Summary}
\label{sectsum}

We present the results of a time-dependent model of a dense photon-dominated
region created after a star has formed inside a dark cloud. We have assumed
for our standard model an homogeneous cloud with density $n=2\times10^4$\cmt\
and gas temperature $T=40$ K, on one side of which impinges an FUV radiation
field $\chi=100$ times the ISRF.

We have found that the more unsaturated and/or simpler hydrocarbons, in
particular \cdh, \cqh, or C$_6$H, have peak fractional abundances in the outer
layers of the PDR, at A$_V\sim$1--2, at times as early as a few $\times 10^4$
yr, while more saturated hydrocarbons, such as CH$_4$, have their peaks deeper
into the cloud at these times, around A$_V\sim5$.  The outer and intermediate
layers reach steady-state rapidly, at times shorter than a few $\times 10^5$
yr, while for the inner layers it takes almost an order of magnitude longer.

The molecule CS and the cyanopolyynes also peak nearer to the surface of the
cloud, at A$_V\sim1.7$--2, at times following a few $\times10^4$ yr, while
HCN, HNC, HCO$^+$, \nthp\ and \nh\ peak deeper into the clouds, from A$_V=5.5$
onwards. This result is very interesting because it reproduces in space a
similar distribution of molecules to the one found in time for a homogeneous
gas-phase model. The external radiation field seems to keep the chemistry
 `young' in the outer layers of the cloud, while only the more shielded slabs
develop a  `late-time' chemistry.

We have run several models with modifications of some of the parameters of the
standard model, varying the intensity of the radiation field, the temperature
and density of the cloud, and the initial distribution of fractional abundances
in the gas. For the three models run with non-standard scaling factors of the
radiation field of 10, $10^3$, and $10^4$, the effects on the peak fractional
abundances of hydrocarbons are not large, typically about an
order-of-magnitude.   At large values of the radiation field, all molecules are
swept out of the outer regions, while only the gas at A$_V> 9-10$ is
completely shielded from variations in the radiation field.  One salient
result of increasing the density is that for hydrocarbons that peak in the
outer layers, the abundance profile when plotted against distance has a
smaller region of large fractional abundance.  Another salient feature is that
the peak fractional abundances of selected hydrocarbons and cyanopolyynes
increase with density in the range 10$^{3}$ - 10$^{6}$ cm$^{-3}$ while other
species show little change or even a decrease. Variations in peak fractional
abundance as temperature is changed tend to be less significant than with
density for temperatures in the range 10-200 K.  We also explored the effects
that different initial fractional abundances have on the results of the model
by running a model that starts with atoms in the gas, finding that the
evolution of fractional abundances in time is very different from the standard
model, although the end result (steady state) is, of course, the same.

In a comparison of the results of our standard model with recent high-angular
resolution observations of the Horsehead nebula, we were able to reproduce,
within the estimated observational uncertainties, the peak fractional
abundance and spatial dependence of \cdh; not so with \cqh\, and \cthd\ which
show much worse agreements at the outermost positions. For the rest of the
molecules, studied at moderate- and high-angular resolution, the agreement to
the observations go from good (for CS and \hcsp) to poor (especially for
\hctn). We have also used additional models to explore how different physical
conditions can improve the agreement with observations. Even if some models
did reproduce better some of the molecular abundances, we could not find a
single model that was able to fit all the observations at the same time.

We also ran two models in which the initial CO abundance was changed
from its  `standard'value of $7.3\times10^{-5}$ with respect to total
H. First, we reduced this value to that measured in TMC1-CP
$(4\times10^{-5})$, while conserving the total elemental abundances of C and O
by increasing the atomic abundances of carbon and oxygen. Secondly, we used
the reduced CO abundance with no attempt to maintain our standard elemental
abundances of C and O. In both instances, little change was seen for
carbon-chain species from the standard model at times near steady state. At
earlier times, however, the first modification produces strong enhancements in
carbon-chain molecules for the inner layers while the second modification
produces smaller diminutions of these abundances that are even smaller than at
steady state. The status of the agreement with observations in the Horsehead
nebula is unaffected by changes in the initial CO abundance, since the changes
affect the inner layers, which have not been observed.

Recently, there have been some new results that might help to solve the
disagreement between observation and theory. First, Rimmer and 
Herbst (in preparation)  have re-calculated the cosmic-ray
ionization rate as a function of depth into clouds. We
found in our models that a different value of $\zeta$ may dramatically change
the abundances of the species. Secondly, there is evidence that calculations
including negative ions may indeed increase the calculated abundances of
neutral carbon chain species. These calculations are currently being done by
R.~Ni Chiumin, Tom Millar, Nanase Harada, Eric Herbst and others. Thirdly,
\citet{parise} have confirmed that the Horsehead Nebula is rather clumpy by
their analysis of deuterated molecules. We intend to use
these new calculations in further work to find out if they are able to
reproduce the abundances of the molecules observed in the Horsehead nebula.

From the currently available observational data, it is difficult to obtain a
definitive age of the Horsehead nebula. We would need more observations,
especially of layers deep into the cloud (A$_V\ga4$), where the abundances of
species such as hydrocarbons and cyanopolyynes are more sensitive to changes
in time.

\section{acknowledgements}

E. H. wishes to thank the National Science Foundation (U. S.) for support of
his research programme in astrochemistry. OM acknowledges the support of the
Na\-tional Science Foundation to the astrochemistry group at Ohio State
Uni\-versity.

%References%%%%%%%%%%%%%%%%%%%%%%%%%%%%%%

\appendix

\section{Calculations made with the newest ratefile osu\_03\_2008}

The calculations discussed in the body of the paper were performed with the
\texttt{osu.2003} network for a variety of reasons.  We have also used the
latest version of this network without molecular carbon chain anions
(\texttt{osu\_03\_2008}) to determine if there are any substantial changes in
the standard version of our model.  Some salient differences found are as
follows:

\begin{enumerate}

\item The abundance of molecular hydrogen is unchanged, and there is very
  little change in the abundance of atomic hydrogen ( $\sim+3$ per cent at
  A$_V\sim2$, $\sim-3$ per cent in the innermost layers). The crossing of the
  H and H$_2$ abundances is at A$_V\sim0.04$. CO abundances are $\sim-5$ per
  cent up to A$_V\sim2$, while C shows the opposite behaviour. Abundances of
  C$^+$ at layers A$_V> 4$ are a factor of $\sim25$ per cent less.

\item Differences between abundances are quite small for most molecules (from
a few percent to a factor of 2) at steady state, especially for the more
abundant hydrocarbons, for which they are typically much less than a factor of
two, except at middle layers (A$_V\sim$4-7), where these differences can go
from 25 per cent less for \cdh\ to a factor of 2 less for \cthd. The main
apparent effect is a slightly lower inner abundance peak for some of these
molecules.  Other neutrals show very small differences, such as HCN
(abundances $20$ per cent larger or lower overall, except a factor of 2 less at
A$_V\sim4$) or H$_2$CO (with almost no changes, less than 5 per cent). \hcop\
shows up and down variations of about 20 per cent at steady state. Some less
abundant hydrocarbons and other low abundance species may show larger
differences at some positions but always at positions where they have very low
abundance.

\item N$_{2}$H$^{+}$ is generally larger with the newer network, with
differences up to a factor of five at some outer layers at steady state, where
the ion is less abundant.  At the inner layers, the differences range from 50
per cent to 20 per cent from early times to steady state. Ammonia does not
show this behavior; its abundance is at most $\sim70$ per cent larger at
middle layers at steady state, but small differences, less than $10$ per cent
in the rest between the two networks at steady state (differences of $\sim40$
per cent at early times).

\item CS and HCS$^{+}$ have lower abundances with the newer network, with
maximum differences a factor of 3-5 for CS and 2-15 for HCS$^+$, especially
for the outer layers, and for $t>5\times10^4$ yr. At the same time, SO and
SO$_2$ show larger abundances for layers A$_V> 6$, of $\sim70$ per cent and a
factor of about 4, respectively, at steady state (with smaller differences at
earlier times).

\item HC$_{3}$N has a larger abundance at steady state, by almost an order of
magnitude, for the outer layers, where it is less abundant, and of $\sim2$ in
the inner layers. At earlier times, the abundances at the inner layers can be
a factor of $\sim5$ larger. The other cyanopolyynes tend to have slightly
lower abundances overall, except a drop at A$_V\sim5$ that can be about an
order of magnitude, but they have very low abundances in both cases.

\item In comparison with the observational results for the Horsehead Nebula,
there are few significant differences.  CS and HCS$^{+}$ would be the most
affected molecules, specially the latter. Nonetheless, the HCS$^+$ abundances
at A$_V\sim2$-4 would still be inside the error bars of the abundance
determined by the observations.

\end{enumerate}

\end{document}